\documentclass[review]{elsarticle}

\usepackage{lineno,hyperref}
\modulolinenumbers[5]

\journal{Astroparticle Physics Journal}
\usepackage{amsmath,amssymb}
\usepackage{epsfig}  
\usepackage{graphicx}               % Standard graphics package  
\usepackage{url}
\usepackage{multirow}
\usepackage{placeins}
\usepackage{float}	
\usepackage{rotating}
\usepackage{subfigure}
\usepackage{blindtext}
\usepackage{url}
\begin{document}

	%\linenumbers

\begin{frontmatter}
	
	\title{{\Large } Identification of Gamma-Rays and Neutrinos from the Cygnus-X Complex
		Considering Radio Gamma Correlation }
	\author[FirstAfilliation]{Mehmet Guenduez}
	\ead{mehmet.guenduez@rub.de}
	\author[FirstAfilliation]{ Julia Becker Tjus}
	\author[FirstAfilliation]{ Bj\"orn Eichmann}%{FIRSTAFF}
	
	\address[FirstAfilliation]{Fakult\"at f\"ur Physik und Astronomie, RAPP Center, TP IV, Ruhr-Universit\"at Bochum, D-44801, Germany}%{FIRSTAFF}
	%{SECONDAFF}
	\author[SecondAfilliation]{Francis Halzen}
	\address[SecondAfilliation]{Department of Physics, IceCube Collaboration, University of Wisconsin, Madison, WI 53706,USA}%{SECONDAFF}
	
	\begin{abstract}
		\noindent
		%Background:
		The Cygnus X region is known as the richest star-forming region within a few kpc and is home to many particle accelerators such as supernova remnants, pulsar wind nebulae or massive star clusters. The abundance of accelerators and the ambient conditions make Cygnus X a natural laboratory for studying the life cycle of cosmic-rays (CRs). This naturally makes the Cygnus X complex a highly interesting source in neutrino astronomy, in particular concerning a possible detection with the IceCube Neutrino Observatory, which has a good view of the northern hemisphere.\\
		%Methods:
		In this paper, we model the multiwavelength spectrum of the Cygnus Cocoon, for the first time using a broad data set from radio, MeV (COMPTEL), GeV (Fermi), TeV (Argo) and 10s of TeV (Milagro) energies. The modeling is performed assuming a leptohadronic model. We solve the steady-state transport equation for leptons and hadrons injected homogeneously in the region and test the role of diffusive transport and energy loss by radiation and interaction.\\
		%Results:
		The result shows that diffusion loss plays a significant role in Cygnus X and always exceeds the advection loss as well as almost all other loss processes. The best-fit parameters we find are a magnetic field of $B=8.9\times10^{-6}$ G, a target density of $N_t=19.4$ cm$^{-3}$, a cosmic ray spectral index of $\alpha=2.37$ and neutral gas distribution over a depth of 116 pc.
		We find that the fit describes the data up to TeV energies well, while the Milagro data are underestimated. 
		This transport model with a broad multiwavelength fit provides a neutrino flux which approaches the sensitivity of IceCube at very high energies ($>$ 50 TeV). In the future, the flux sensitivity of IceCube will be improved. With this rather pessimistic model, leaving out the influence of possible strong, high-energy point sources, we already expect the flux in the Cygnus X region to suffice for IceCube to measure a significant neutrino flux in the next decade.

	\end{abstract}
	
\end{frontmatter}

%\linenumbers

\section{Introduction}
\label{Introduction}
\noindent One of the main unsolved problems in astroparticle physics is based upon the origin of high energetic cosmic rays (CRs). Although more than 100 years have passed since Victor Hess discovered cosmic rays (\cite{VictorHess}), the search for a reliable answer still continues.
An insight into the acceleration mechanism which causes the energy gain plays an important role in that regard. Active Galactic Nuclei (AGN), Supernova Remnants (SNRs), Gamma-Ray Bursts (GRBs) and pulsars wind nebulae (PWN) are showing great promise for being accelerators of CRs (i.e. \cite{HighEnergyAstrophysics}). \\
CRs are deflected by magnetic fields and interact with the ambient medium (\cite{Ginzburg}). The deflection makes it harder to locate the source of CR, whereas for high energy photons and neutrinos this problem does not occur. In fact,  high energy (TeV) photons, but especially astrophysical neutrinos, can point to the direction of the source (\cite{Julia1}). These neutrinos also allow us to draw conclusions about the hadronic particles themselves since their generation process is based on the interaction between hadronic particles and the ambient medium by producing pions.\\
%CRs generate secondary particles (neutrinos, muons, mesons, electrons, electromagnetic radiation) in the GeV range and above as a result of leptonic or hadronic processes. %Secondaries from galactic CRs take the title PeVatrons and should be observable by multiple detectors such as Fermi, Argo or IceCube. 
Galactic CR accelerators are called PeVatrons as they accelerate up to the knee, i.e. PeV energies. The associated particles should be observable by multiple observatories such as Fermi, Argo or IceCube. Even if these detectors provide accurate data, it is necessary to identify the relevant radiation processes to give a realistic interpretation of the experimental data. In particular, the origin of $\gamma$-rays can be explained by several processes.\\
The brightest diffuse $\gamma$-ray emission in the northern hemisphere is detected from the Cygnus X complex. It reeveals that this astrophysical region is rich on cosmic-ray accelerators (see e.g.\ \cite{ARGOCyg}). Cygnus X can expose the secret behind the acceleration mechanism of the CRs because of the short distance between the Earth and Cygnus X and the content of well-studied sources (such as J2032.2+4126, J2021.0 + 3651, J2021.5+4026 or J2030.0+36542). In order to comprise them and potential accelerators such as SNR and PWN, our region of interest (ROI) includes these known sources within a radius of $3.14 \deg$ centered in Cygnus X and a solid angle of $32 \deg ^2$, respectively, as this part of Cygnus includes the dominant part of the high-energy emission.\\
In previous works \cite{Francis} and \cite{IdentifyingPevatrons}, the neutrino flux from Cygnus X was calculated by simple approximations and assuming parameters which are averaged over the Galaxy, e.g. the magnetic field strength ($B$=1 $\mu$G ). Today, it is known that these parameters could deviate significantly from the used values, indicated by different astrophysical observations. In our model, we use parameters like the magnetic field and the column depth as free parameters and determine them via a best-fit scenario.\\
Recently, \cite{TovaPaper} investigated Cygnus X more extensively by assuming that the CR spectrum observed at Earth is also a representative for the Cygnus X and by adding emission from the Cocoon as well as from point sources separately to the diffuse emission modeled in their work. Moreover, continuous momentum loss and losses due to advection and diffusion were considered.
All calculations were carried out for 5 deg $\times$ 5 deg region which is subdivided in 0.25 deg $\times$ 0.25 deg, and considered observation data from 150 MeV (Fermi) up to 16 TeV (Milagro).\\
In this paper, we add information from radio wavelength and also take into account the COMPTEL-detected 10-MeV signal. This broad energy range gives strong constraints on the possible leptonic (synchrotron, non-thermal Bremsstrahlung, inverse Compton) and hadronic (pion production) processes in Cygnus X. In doing so, the transport and the loss mechanisms in Cygnus X can be investigated, such that the resulting neutrino spectrum is derived.\\
The primary requirement to generate high energetic CRs is an appropriate accelerator, which in Cygnus X is thought to be  PWN or SNRs.\\
\noindent In the same regard, different losses will be considered:
\begin{enumerate}
	\item Continuous momentum losses by: Synchrotron, Inverse Compton, non-thermal Bremsstrahlung, ionization and hadronic pion production.
	\item Catastrophic losses by advection and diffusion.
\end{enumerate} 
Per definition, continuous momentum losses conserve the total number of the particles in Cygnus X, whereas catastrophic losses do not. This means that particles escape the region of interest due to diffusion or advection.
In the same vein, both the flattening of the hadronic pion production for energies greater than 200 GeV as well as the different cooling behaviors of electrons and protons will be considered (\cite{Kelner}).\\
%The diffusion timescale seems to be much shorter  than the advection timescale and therefore the diffusion might more dominant.
% are there 4 or 6pulsars? 
%%%approach step:
Since the exact particle accelerator is not known, the acceleration mechanism of CRs from that region will remain unspecified. Therefore, a CR emission from a non-thermal electron-proton plasma with a power-law in momentum will be used.  This work will rely on the mathematically convenient description by assuming a spatially homogeneous and spherically symmetric CR density distribution in Cygnus X since the region is very complex and small inhomogeneities vanish at a larger scale.\\
Nevertheless, the rigidity difference between electrons and protons will be considered.
%\footnote{Here, messenger particles denote particles which carry directional information from their sources. In the following, the usage will be restricted to photons and neutrinos.} 
Also, the messenger particle from secondaries of CRs will be considered to find confirmation indirectly for the proposed model by examining experimental data.

\section{Cygnus X}
\begin{figure}[H]
	\centering	\includegraphics[width=0.8\linewidth]{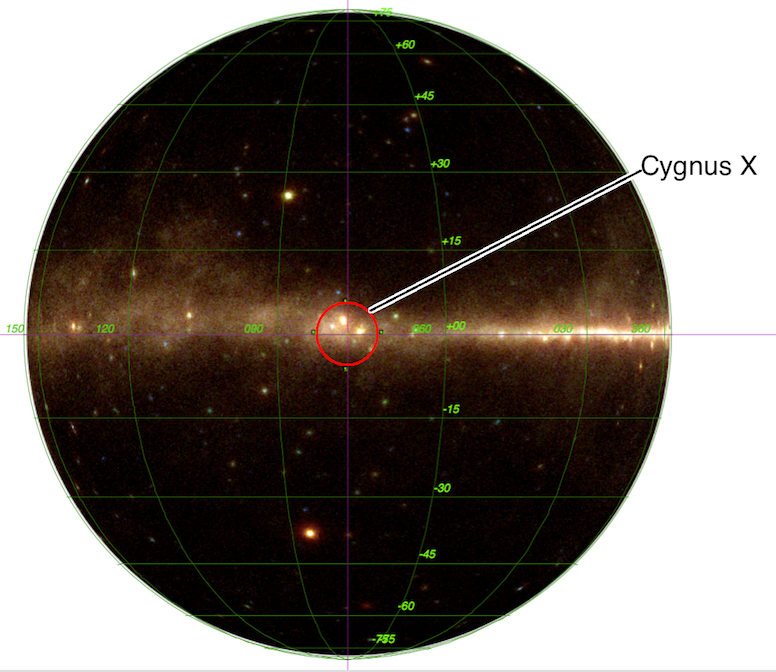}
	\caption[Fermi's color map]{Fermi's color map from -60$^{\circ}$ - +60$^{\circ}$ in the vertical plane and from 150$^{\circ}$ - 300$^{\circ}$ in the horizontal plane, which is distributed by Skyview HEASARC - HEALPixed by CDS; the map was edited with Aladin v9.0. The red color in the map denotes photons in the energy range 0.3-1 GeV, green 1-3 GeV and blue 3-300 GeV (\cite{AladinFermi}).}
	\label{fig:fermimap}
\end{figure}

\begin{figure}[H]
	\centering
	\includegraphics[width=0.8\linewidth]{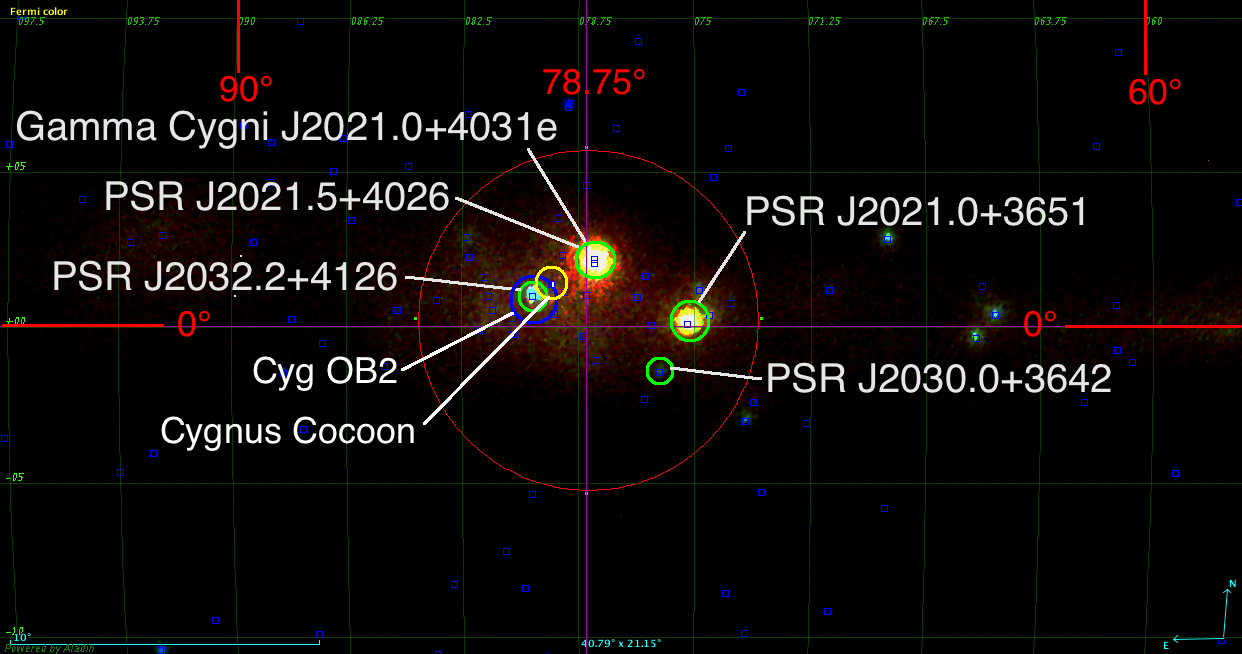}
	\caption[Cygnus X from Fermi's color map]{Fermi's color map of Cygnus X (red cycle), which is distributed by Skyview HEASARC - HEALPixed by CDS; the map was edited with Aladin v9.0. The scales are given by galactic coordinates. Cygnus X is represented here by the big red cycle, pulsars abbreviated with "PSR" by green cycles, OB2  association by a blue cycle and Cygnus Cocoon by a yellow one. All sources identified by Fermi 3FGL are represented by blue quads.}
	\label{fig:fermimapzoom2b}
\end{figure}

\noindent Cygnus X is a part of the largest star-forming region of the constellation Cygnus in the northern galactic plane, which is located in the galactic local spiral arm, more precisely at galactic longitudes between $70^{\circ}$ and $90^{\circ}$ and $4^{\circ}$  and $8^{\circ}$  below and above the plane (see figure \ref{fig:fermimap}) (\cite{CygnusByFermi2}). It is one of the most structurally complex regions in the galactic plane. Moreover, it is formed by a massive molecular cloud complexes. This property is important for CR formation and characteristics. Nevertheless, as indicated in \cite{CygnusByFermi} the CR population is similar to the local interstellar space. 
\noindent There are many reasons why Cygnus X is an excellent region to investigate the origin of CRs:
\begin{itemize}
	\item The emission is observable from radio to high-energy gamma-ray frequencies (\cite{GammaSourcesCygnus}), whereby in the energy range from GeV up to TeV Cygnus X has the brightest emission in the northern hemisphere (\cite{BrightestSourceCygnus}). Moreover, many other gamma-ray sources exist in that region.
	\item  It contains sources which accelerate particles at least up to 100 TeV  (\cite{Milagro}).
	\item Many potential accelerators such as supernova remnants\footnote{For example $\gamma$ Cygni J2021.0+4031e, which Milagro also detected at very high energies (\cite{MilagroGammaCygni}).}, pulsar wind nebulae\footnote{List of pulsars in the region of interest: J2032.2+4126, J2021.0 + 3651, J2021.5+4026, J2030.0+36542.} and Wolf-Rayet (WR) stars or  OB associations (Cyg OB2, Cyg OB1) can be found. Many of these constituents are pictured in figure \ref{fig:fermimapzoom2b}.
	Here, it is important to mention that approximately 20\% of CRs are produced in WR stars nearly $10^5$ years before they become accelerated (\cite{CygnusByFermi2}). These stars are a phase of OB stars, which appear in Cygnus X as clusters (OB associations).
	\item Most of the objects are at a distance of 1.4 kpc.
	\item  It consists of H$_{\text{II}}$ regions (\cite{IdentificationTeVCygnusCocoon}).
	
\end{itemize}
\noindent All of these characteristics make Cygnus X a suitable natural laboratory for the astronomer to look beyond the usually constrained view.
%Figure \ref{fig:cocooncygnusdiffusivefluxpaper}  exhibits the column density of H$_{\text{I}}$ by means of galactic coordinates.
%\begin{figure}[H]
%	\centering
%\includegraphics[width=1.0\linewidth]{Pictures/CocoonCygnusDiffusiveFluxPaper}
%	\caption[Column density of Cygnus X]{H$_{\text{I}}$ column density as a function of the galactic coordinates in Cygnus X in the local Spur (left figure) and outer Galaxy (right figure) with a spin temperature of 250 K; the color scale is given in 10$^{20}$ atoms/cm$^2$ \cite{CygnusByFermi}.}
%	\label{fig:cocooncygnusdiffusivefluxpaper}
%\end{figure}
%\begin{figure}[H]
%	\centering
%	\includegraphics[width=1.0\linewidth]{Pictures/SciencePaper1}
%	\caption[Photon count map from 10-100 GeV of Cygnus X]{Relative photon count map from 10 - 100 GeV by Fermi; "A" pictures the total emission, "B" the total emission after subtracting all interstellar background and all known sources except $\gamma$ Cygni and "C" the total emission after further removal of $\gamma$ Cygni \cite{Sciencepaper}.}
%	\label{fig:sciencepaper1}
%\end{figure}
%The emissivity of the atomic gas within Cygnus X seems to be consistent with the interstellar medium \cite{FermiData}.\\
\noindent The supernova remnant (SNR), $\gamma$ Cygni, was firstly investigated using Fermi data, which provide information about the interstellar background by subtracting the radiation from $\gamma$ Cygni.\\Moreover, Cygnus X has a Cocoon where freshly accelerated CR can be found, and the emission exceeds 100 GeV.  The SNR $\gamma$ Cygni, which is located in the Cocoon, could cause the acceleration of protons even up to 80-300 TeV and electrons up to 6-30 TeV. %Figure \ref{fig:sciencepaper1} shows the influence of the interstellar background emission but also of $\gamma$ Cygni.\\
The accelerated particles could fill the whole Cocoon if it is assumed that the primary transport mechanism is diffusion. On the other hand, advection could dominate the transport mechanism, if an anisotropic emission from $\gamma$ Cygni was observed (\cite{CygnusByFermi2}). However, there is no proof for this scenario yet.\\The Cocoon can give hints about the transport mechanism and escape of CRs from their source.
\noindent In the model built in this work, the influence of diffusion and advection in Cygnus X can be investigated. Thus, at the very least a suggestion of the role of $\gamma$ Cygni in the Cocoon can be given as our ROI includes these objects.\\
In order to properly model the CR interactions, the column depth needs to be known. Following \cite{CygnusByFermi}, we will use $7\times10^{21}$ atoms/cm$^2$ for ROI.\\
\subsection{Local distribution in radio and gamma range}
\noindent In our model, we assumed a spatially homogeneous injection of accelerated CRs. It is important to investigate the local distribution to see the reliability of this assumption.
\begin{figure}[H]
	\centering
	\includegraphics[width=0.7\linewidth]{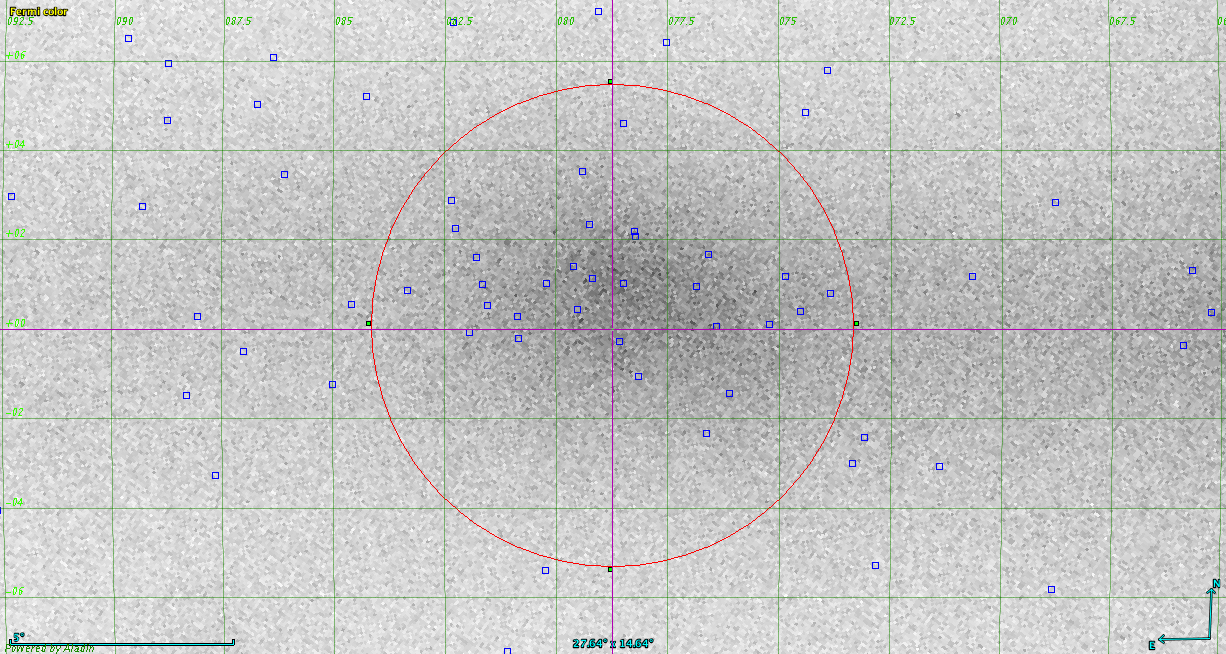}
	\caption{Fermi map band 1: 30-300 MeV}
	\label{fig:fermimapzoom11}
\end{figure}
\begin{figure}[H]
	\centering
	\includegraphics[width=0.7\linewidth]{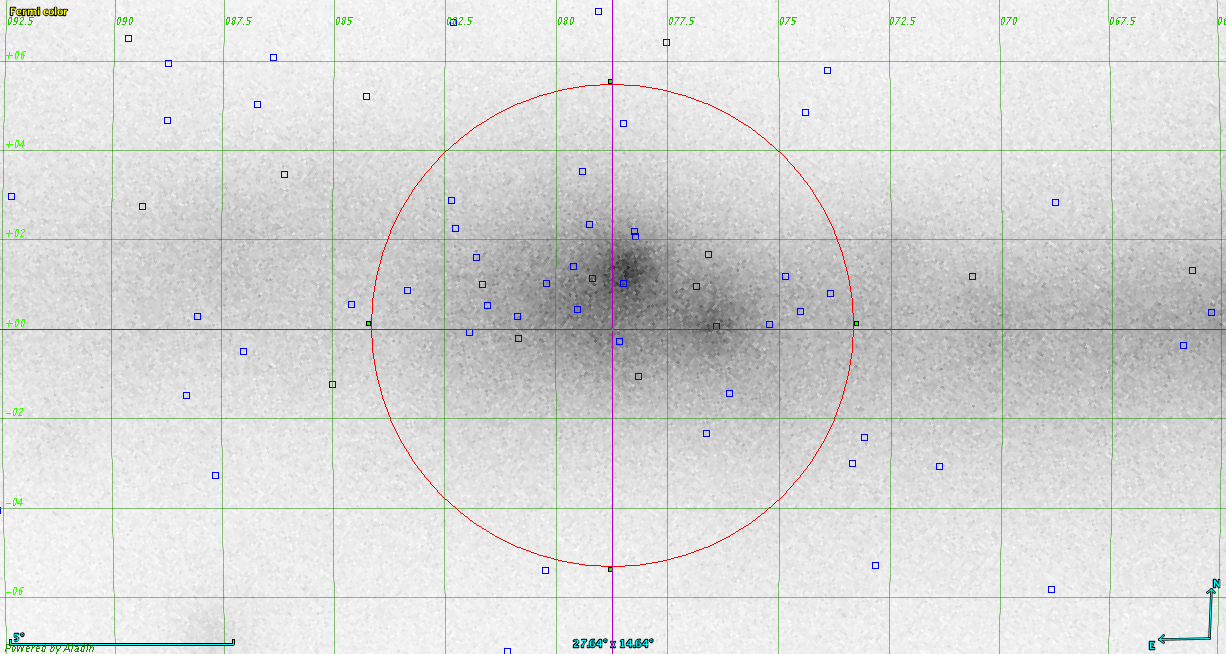}
	\caption{Fermi map band 2: 0.3-1 GeV}
\end{figure}
\begin{figure}[H]
	\centering
	\includegraphics[width=0.7\linewidth]{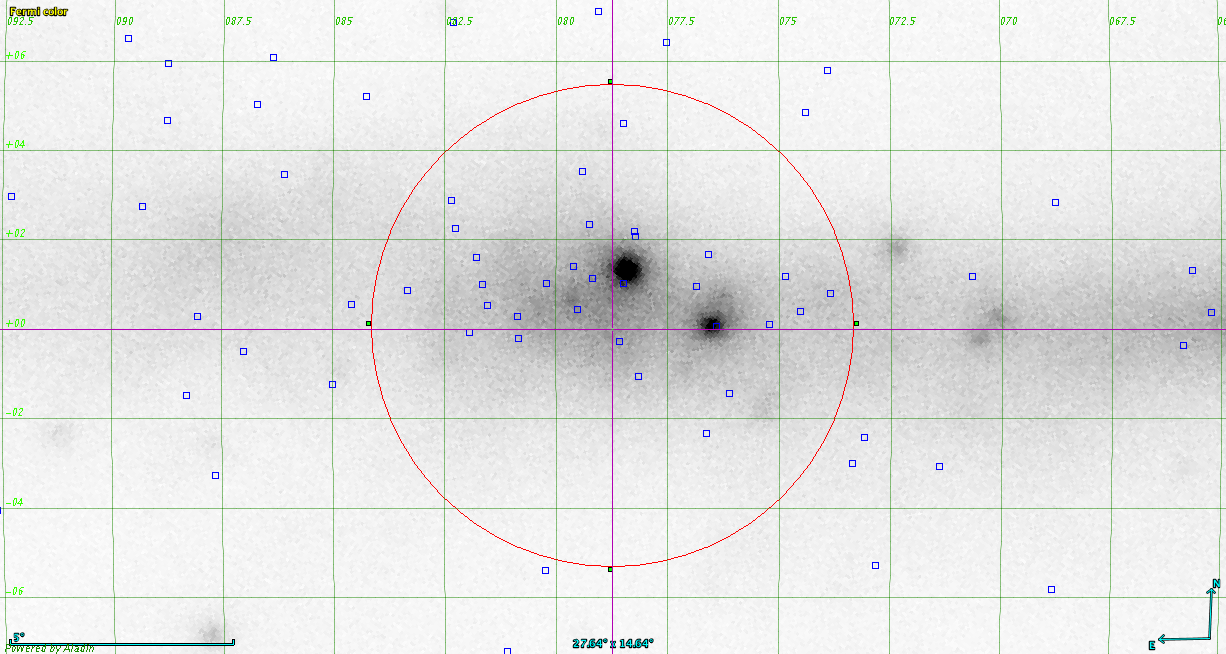}
	\caption{Fermi map band 3: 1-3 GeV}
\end{figure}
\begin{figure}[H]
	\centering
	\includegraphics[width=0.7\linewidth]{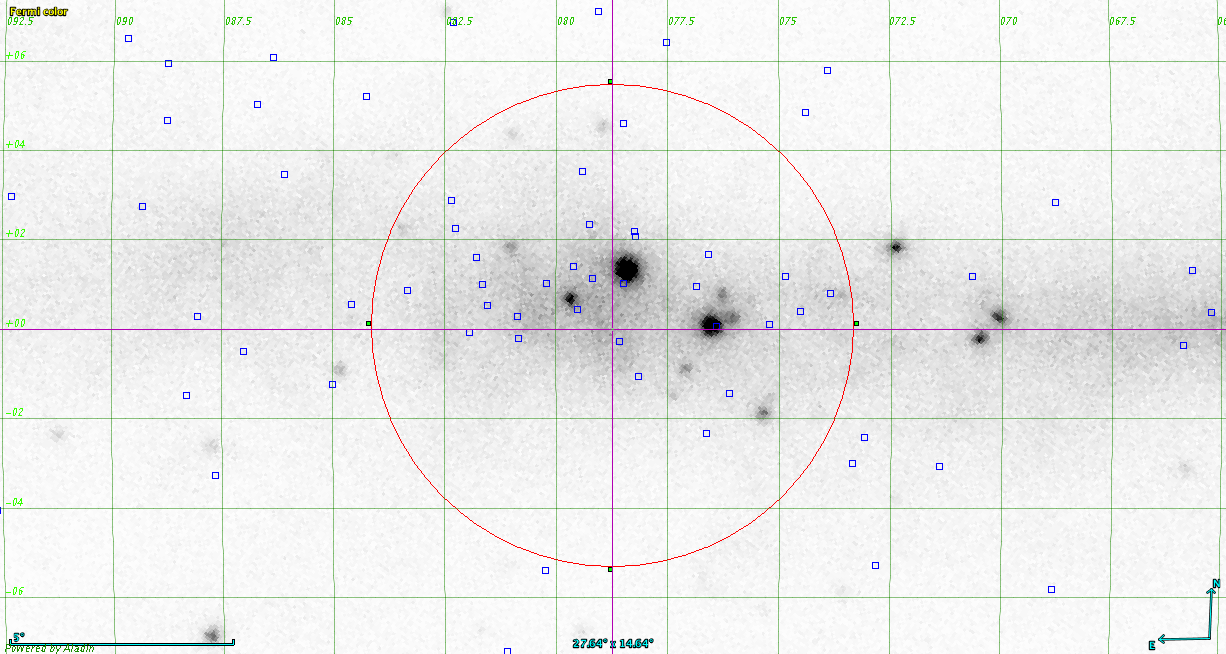}
	\caption{Fermi map band 4: 3-10 GeV}
\end{figure}
\begin{figure}[H]
	\centering
	\includegraphics[width=0.7\linewidth]{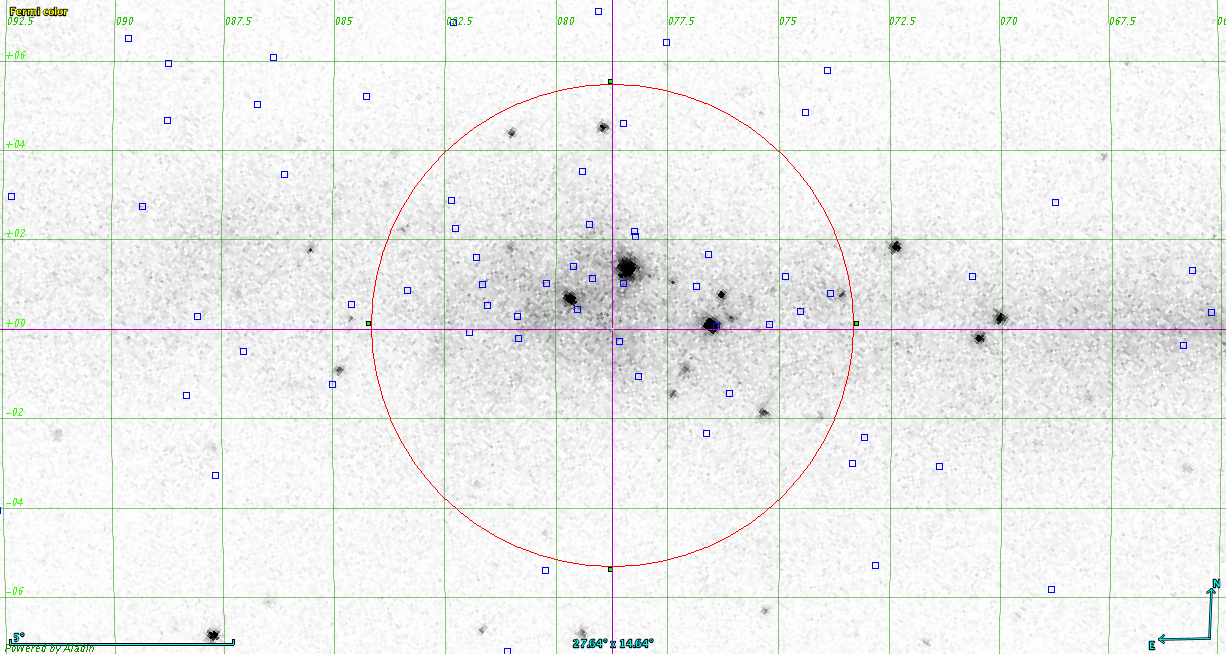}
	\caption{Fermi map band 5: 3-300 GeV}
\end{figure}
\begin{figure}[H]
	\centering
	\includegraphics[width=0.7\linewidth]{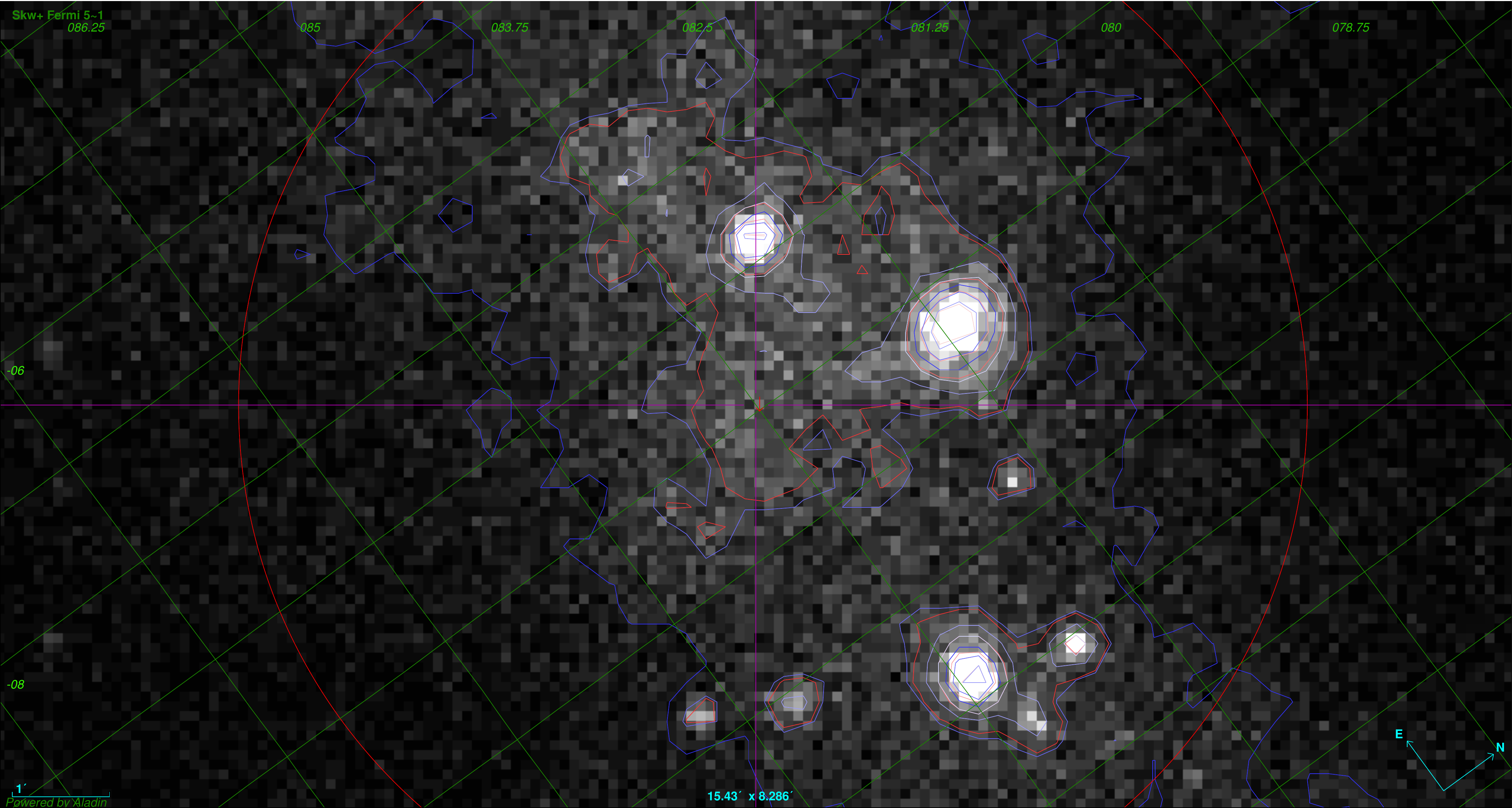}
	\caption{Fermi map band 5 with contours by reducing the background emission}
	\label{fig:fermimapzoom12}
\end{figure}
\begin{figure}[H]
	\centering
	\includegraphics[width=1.0\linewidth]{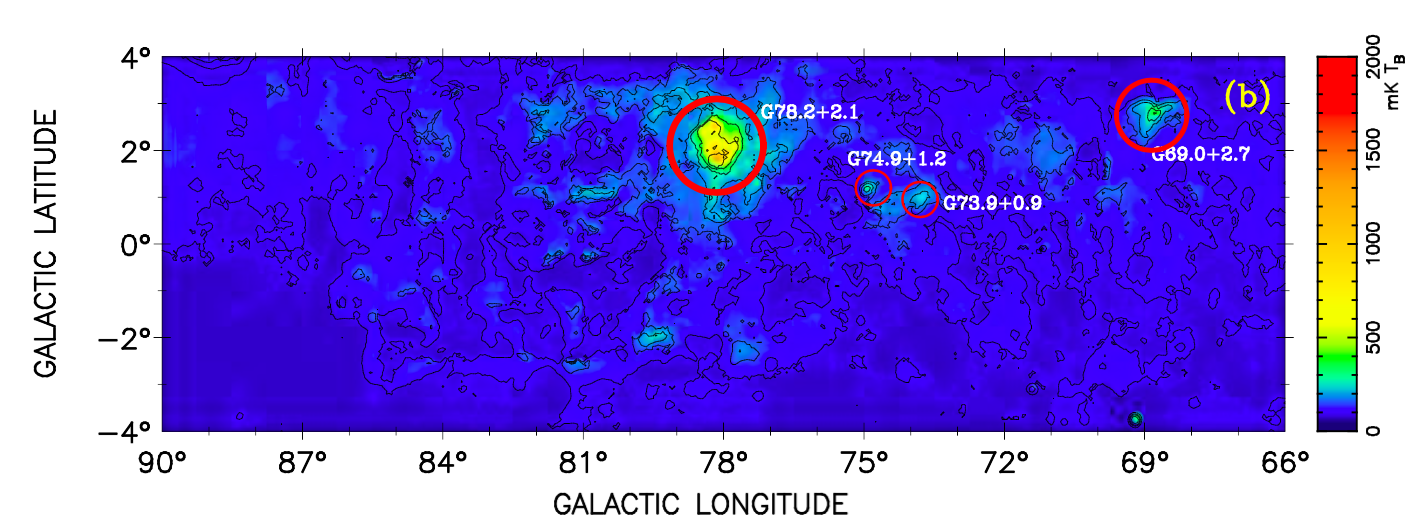}
	\caption[Non-thermal intensity map at 4800 MHz]{Decomposed non-thermal emission from the total intensity in Cygnus X at 4800 MHz with an angular resolution of 9'.5 (\cite{thermalNonThermal1}).}
	\label{fig:non_thermal}
\end{figure}
\noindent Figures \ref{fig:fermimapzoom11} -\ref{fig:fermimapzoom12}
which are extracted from \cite{AladinFermi} for Fermi bands represent a photon count map and serve here as visualization of the structure of Cygnus X dependent on the energy. The density of counts is anti-proportional to the brightness. However, for our calculations we will use directly data from Fermi. \\ For Fermi band 1, the structure is to some extent compatible with our assumptions of a spatially homogeneous and spherical symmetric distribution of CRs in Cygnus X. This agreement deteriorates with higher energies. Therefore, a stronger agreement for lower $\gamma$-ray energies can be expected than for higher energies.\newpage
\subsection*{Cygnus X-1}
\noindent Cygnus X-1 is a binary X-ray system similar to Cygnus X-3. It is also a microquasar,  which could produce PeV $\gamma$-rays in Cygnus X (\cite{Schlicki}). A microquasar is a binary star system with a stellar black hole or a neutron star. Cygnus X-1 contains a black hole with nearly 15 $M_{\odot}$. The main star is a blue giant (HD 226868), and the constellation seems to have been in existence for almost $5\times10^6$ years (\cite{UniverseinGammaRays}).\\
This constellation has a soft and hard state and most of the time remains in the latter. This non-thermal component is thought to be caused in an optically thin and hot corona by thermal Comptonization of disk photons (\cite{CygnusX1}). During the latter, the energy of the constellation is in the power-law component. Therefore, especially at 10 MeV, a purely non-thermal emission will be expected.
\section{Introduction of the model}
\label{Model} 
\begin{figure}[H]
	\centering
	\subfigure{\includegraphics[width=0.8\linewidth]{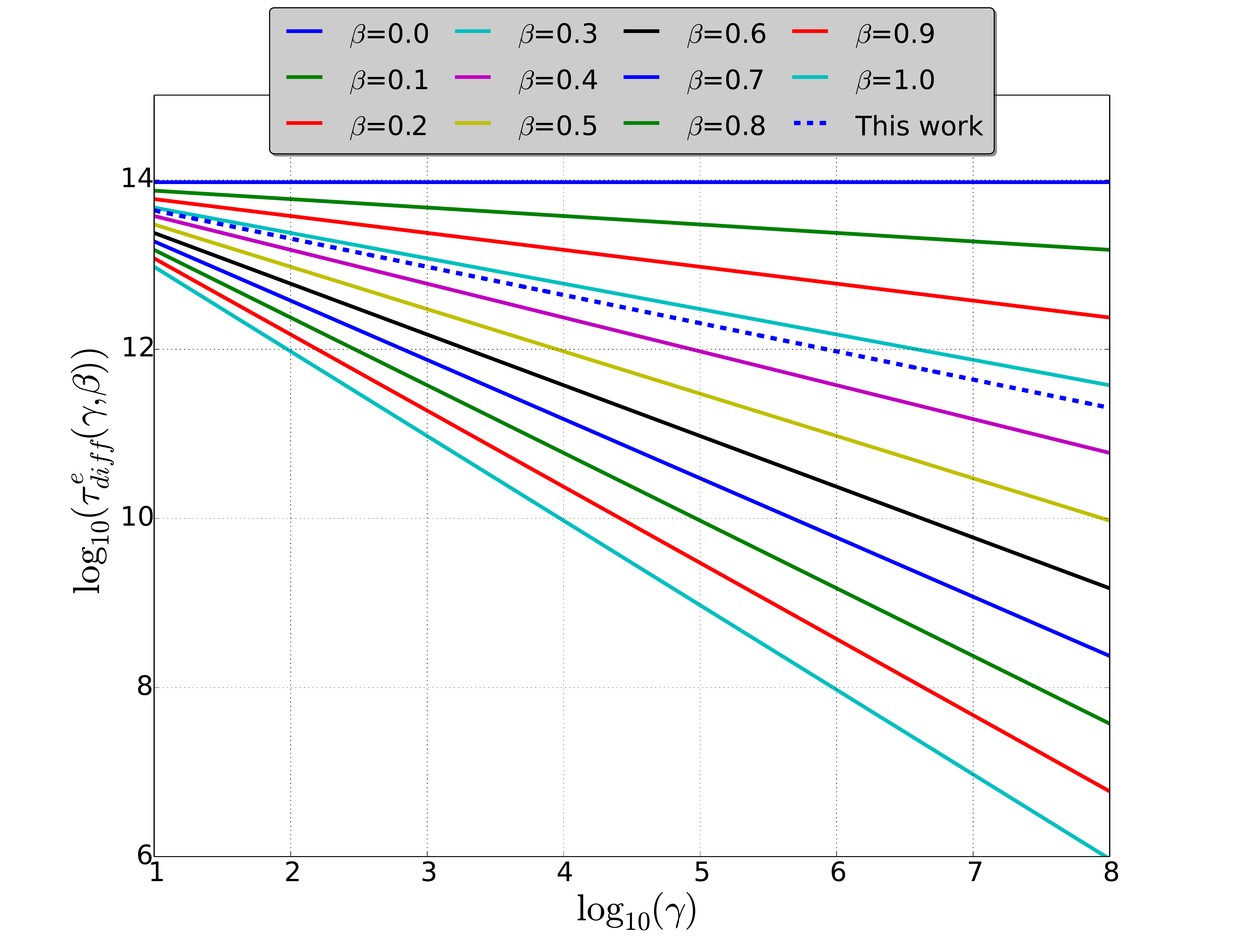}}
	\caption[Electron diffusion timescale]{Electron diffusion timescale $\tau_{diff}^{e}(\gamma,\beta)$ as a function of the related spectral index $\beta$ and Lorentz factor $\gamma$.}
	\label{TimescaleE}
\end{figure}

\begin{figure}[H]
	\centering
	\subfigure{\includegraphics[width=0.8\linewidth]{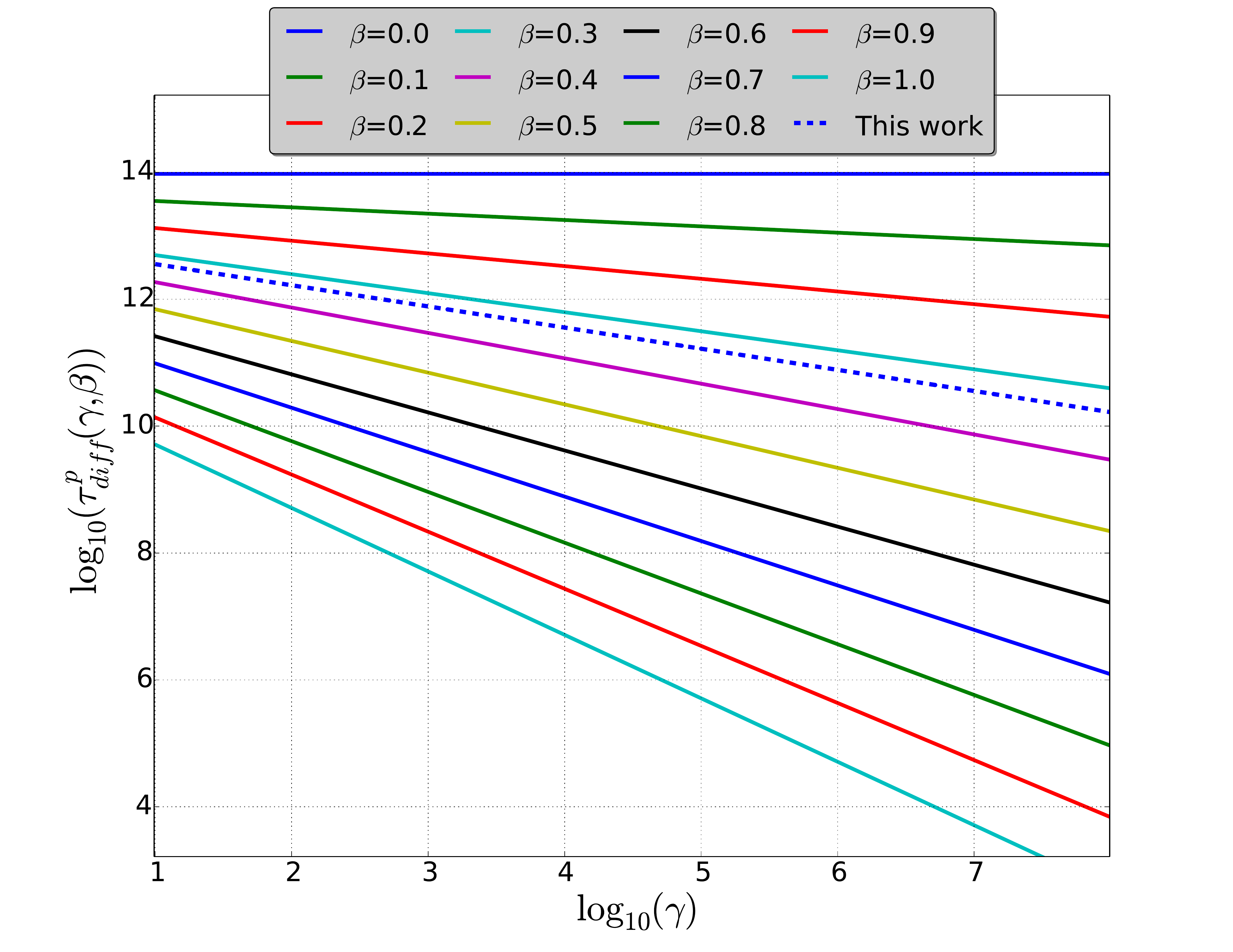}}
	\caption[Proton diffusion timescale]{Proton diffusion timescale $\tau_{diff}^{p}(\gamma,\beta)$ as a function of the related spectral index $\beta$ and the Lorentz factor $\gamma$.}
	\label{Timescale}
\end{figure}

\begin{figure}[H]
	\centering
	\includegraphics[width=0.8\linewidth]{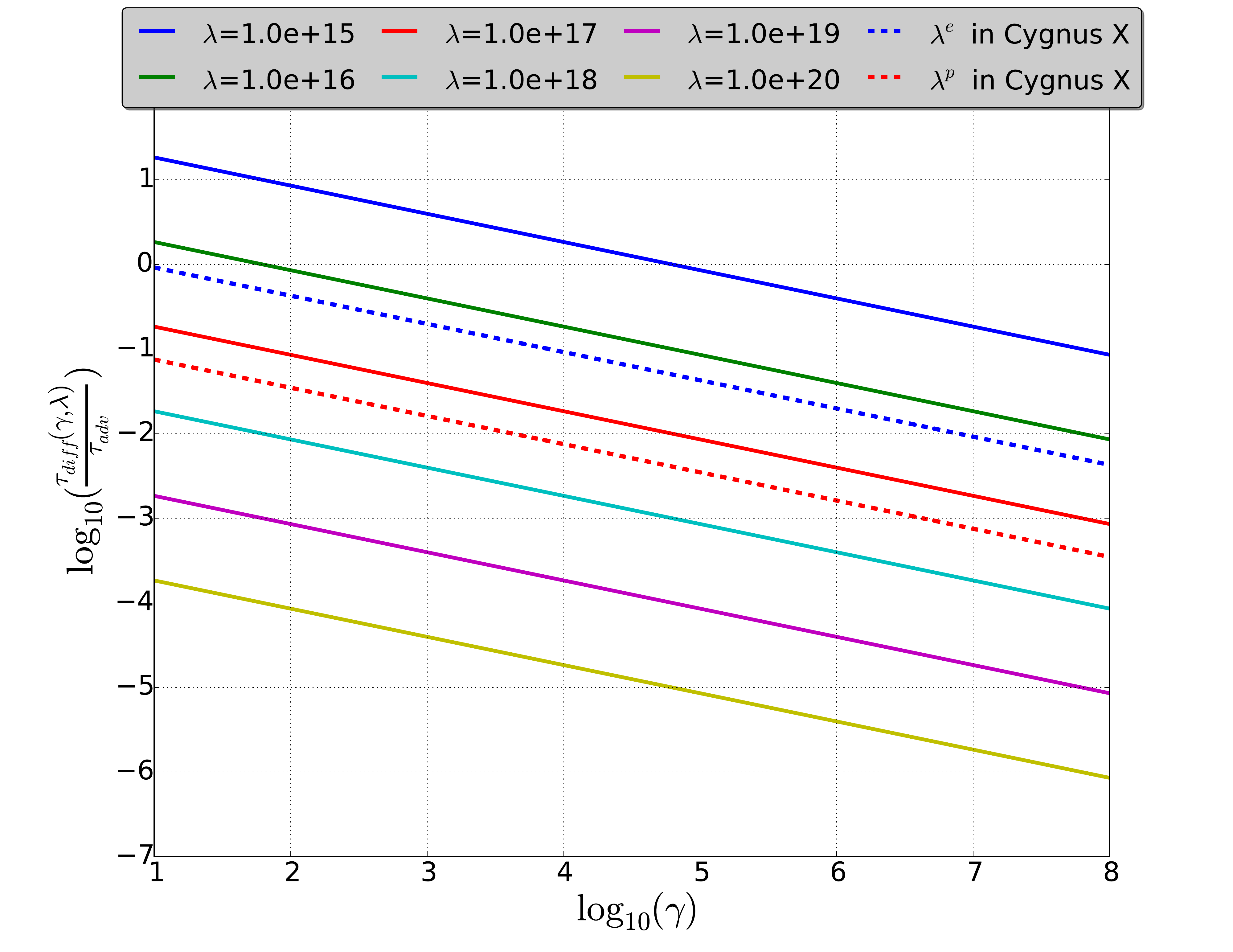}
	\caption[Ratio of diffusion and advection timescale]{The ratio of diffusion and advection timescale as a function of the diffusion length coefficient $\lambda$ and Lorentz factor $\gamma$. Additionally, the dashed line for the proton (red) and electron (blue) diffusion length exhibits the ratio in Cygnus X.}
	\label{fig:timescaleleelectron}
\end{figure}

\noindent The dynamics in a star-forming region can be very complicated a fortiori in the Cygnus X complex. It is, therefore, necessary to work by simplifying assumptions, which are reasonable and favorable to the relevant conditions.\\
The ability to describe particle transport phenomena is indispensable for predicting processes in star-forming regions. The massive molecular clouds in Cygnus X demand a reliable transport mechanism. Beyond the common continuous momentum losses, the catastrophic losses due to advection and diffusion of particles will also be considered.
\\
\noindent Considering the generic state transport equation, the transport equation for a plasma with a differential particle density $n$ yields:
\begin{equation}
\begin{split}
&\underbrace{\frac{\partial n(\textbf{x},\textbf{v},t)}{\partial t}}_{\substack{\text{Storage}}} -\underbrace{\nabla_{\textbf{v}}(F(\textbf{v})\,n(\textbf{x},\textbf{v},t)))}_{\substack{\text{Continuous Loss}}}+ \underbrace{\nabla(\textbf{v}\,n(\textbf{x},\textbf{v},t))}_{\substack{\text{Advection}}}\\&+\underbrace{\nabla(D\,\nabla n(\textbf{x},\textbf{v},t) )}_{\substack
	{\text{Diffusion}}}=\underbrace{q(\textbf{x},\textbf{v},t)}_{\substack{\text{Generation}}}\, .
\end{split}
\label{generalTrEq}
\end{equation}
Here, $\textbf{v}$ is the advection velocity, in consideration it represents the galactic wind velocity. The diffusion tensor is approximated by the scalar $D$.\\
The first term describes the storage; the second term in eq.(\ref{generalTrEq}) the continuous losses in momentum; the third, the catastrophic losses due to advection in magnetic fields; the fourth, the catastrophic diffusion losses; and the last term, the source rate which obeys a power-law. \\ \\
This work is developed to describe an emission from an isotropic and spatially homogeneous part of the Cygnus region in its steady state. Additionally, it assumes an isotropic diffusion of the particles within Cygnus X. 
This assumption is reasonable since an extended region with a diameter of 77 pc will be considered, emission outside this region is negligible, the region is very complex and small inhomogeneities vanishes at larger scales\footnote{Small inhomogeneities vanish especially at larger scales than the gyro-radius.}\\
In our model, we follow a general leaky box approach in which the cosmic-ray density does not depend on spatial coordinates and is characterized by some average values \cite{Ginzburg}. This general scheme is well-established with previous analytical solutions for different scenarios given in \cite{Schlicki} and references therein. Here, we particularly follow the model of \cite{Bjoern}, who meets the requirements we apply to our source region, i.e. a homogeneous steady-state CR sea with a power-law-injection, continuous momentum as well as diffusion and advection loss.\\
With these assumptions, from eq.(\ref{generalTrEq}) the following equation is obtained (\cite{Schlicki, Bjoern}) as the steady state transport equation:
\begin{equation}
0=\underbrace{\frac{\partial}{\partial \gamma}\left( \Gamma_{e,p}n_{e,p}(\gamma)\right) }_{\substack{Continuous \\momentum\ loss}}-\underbrace{\frac{n_{e,p}(\gamma)}{\tau^{e,p}_{diff}(\gamma)}}_{\substack{Diffusion\\ Loss}}-\underbrace{\frac{n_{e,p}(\gamma)}{\tau^{e,p}_{adv}(\gamma)}}_{\substack{Advection\\ Loss}}+\underbrace{q_{e,p}(\gamma)}_{\substack{CR\ source\\ rate}}\, .
\label{TransportEq}
\end{equation}
Here, $\gamma$ represents the Lorentz factor, $\Gamma$ is the term which includes the continuous loss and $\tau_{adv}$ the advection loss timescale.
Using the assumption mentioned above for catastrophic losses the diffusion coefficient for electrons or protons can be approximated by:
\begin{equation}
D_{e,p}(\gamma)\simeq\frac{c\,\lambda_{e,p}\,\gamma^{\beta}}{3}\ .
\label{diffsuionCoeff} 
\end{equation}
The diffusion timescale  $\tau_{diff}^{e,p}$ can then be approximated (\cite{Bjoern}) by:
\begin{equation}
\tau_{diff}^{e,p}(\gamma)\simeq \frac{R^2}{3D(\gamma)}\simeq \frac{R^2}{c\,\lambda_{e,p}}\gamma^{-\beta}\ .
\label{sigmaDiff}
\end{equation}
Here, $\lambda_{e,p}\, \gamma^{\beta}$ denotes the related diffusion length or the mean free path, $\beta$ the related spectral index of the diffusion coefficient, $R$ the radius of the considered region and $c$ the speed of light. The diffusion in the astrophysical context does not describe deflection by collision but interaction with local magnetic fields. As more deflections and interaction force the particles to be spread in the region, the diffusion timescale equals the particle escape timescale and is a quantity for particle conservation.\\
Observation of C/O nuclei spectra shows that the scalar diffusion coefficient index (see eq.(\ref{diffsuionCoeff})) in the Galaxy is given by $\beta\approx 0.5$ (\cite{DiffGalaxyRatio}). Though, to accord with the observation data, this work uses a Kolmogorov-spectrum referring to $\beta=1/3$.\\
The rigidity difference between electrons and protons will consistently be taken into account. This has an effect on the maximum energy and ratio of momentum loss through various phenomena. In this manner, electrons can reach higher energies faster than protons because of their mass difference. Since particles at higher energies are more conducive to continuous momentum loss, it follows that the latter is more prevalent among electrons. The loss mechanisms are not the same for electrons and protons due to their characteristics as leptons and baryons.\\
For protons in Cygnus X, the diffusion length factor is supposed to be $\lambda_p=2.5\times10^{17}$ cm. Due to the mass dependence of the Larmor radius the electron mean free path can be calculated as follows:
\begin{equation}
\lambda_e=\left(\frac{m_p}{m_e}\right) ^{-\beta}\lambda_p\, .
\end{equation}
The influence of the spectral index  $\beta$ on the diffusion timescale can be seen in figure \ref{Timescale} and  \ref{TimescaleE}. This index describes the energy dependence of the diffusion coefficient.
%, which has to be calculated for different energy ranges separately, e.g. for electrons with bigger and smaller 2 GeV \cite{Schlicki. Miller}. However, the energy dependency and thus the diffusion index $\beta$ depends primarily on the turbulence spectrum, which can not be determined for far astronomical objects \cite{Schlicki}. Therefore a best fit index of $\beta=0.25$ will be used. This assumption is necessary for the model in this work, as it hast to reproduce the measured data  and this value seems best to describe the total flux at 10 MeV.
\noindent The timescale becomes smaller at higher energies, and the $\beta$ reinforces this behavior with larger value. However, the influence of diffusion for electrons and protons in Cygnus X can be seen by considering the blue dashed line in figures \ref{TimescaleE} and \ref{Timescale}. Since the advection timescale is constant, the influence of the diffusion length factor $\lambda$ on the diffusion timescale in Cygnus X can also be seen in figure \ref{fig:timescaleleelectron}.
\noindent The galactic wind speed can be used to determine the advection timescale according to
\begin{equation}
\tau^{e}_{adv}=\tau^{p}_{adv}=\tau_{adv}\simeq\frac{R}{v_{wind}}\, .
\label{sigmaAdv}
\end{equation}
For Cygnus X the advection velocity is assumed to be $v_{adv}\simeq50$ km/s as we now it from the Galactic Disk \cite{WindVel}, which is comparable to the Alfvén speed of the CRs.
Using the relation in eq.(\ref{sigmaDiff}) and eq.(\ref{sigmaAdv}) the diffusion and advection loss for Cygnus X can be compared by considering the ratio of both timescales. 
\noindent Figure \ref{fig:timescaleleelectron} shows the ratio as a function of the diffusion length coefficient $\lambda$ and the Lorentz factor. The ratio for relativistic electrons and protons in Cygnus X can be found by considering the advection timescale  $\tau_{adv}=4.752\times10^{13}$ s and the diffusion timescale $\tau_{diff}^e=6.67\times10^{13}\times\gamma^{-\beta}$ s and $\tau_{diff}^p=3.26\times10^{9}\times\gamma^{-\beta}$ s.
\noindent Concerning catastrophic losses, relativistic particles in Cygnus X are more subject to diffusion loss than to advection loss, as the diffusion timescale is much shorter.\\
The timescale for continuous loss be approximated by %\todo{approximately or exactly?}
\begin{equation}
\tau_{con}^{e,p}= \frac{\gamma}{\Gamma_{e,p}}\, .
\end{equation}
In the same manner, the timescale dependency for electron and proton continuous loss can be calculated when the momentum loss rate $\Gamma_{e,p}$ is known.
\subsection{Relation between electrons and protons}
\label{RelationEP}
\begin{figure}[H]
	\centering
	\includegraphics[width=0.9\linewidth]{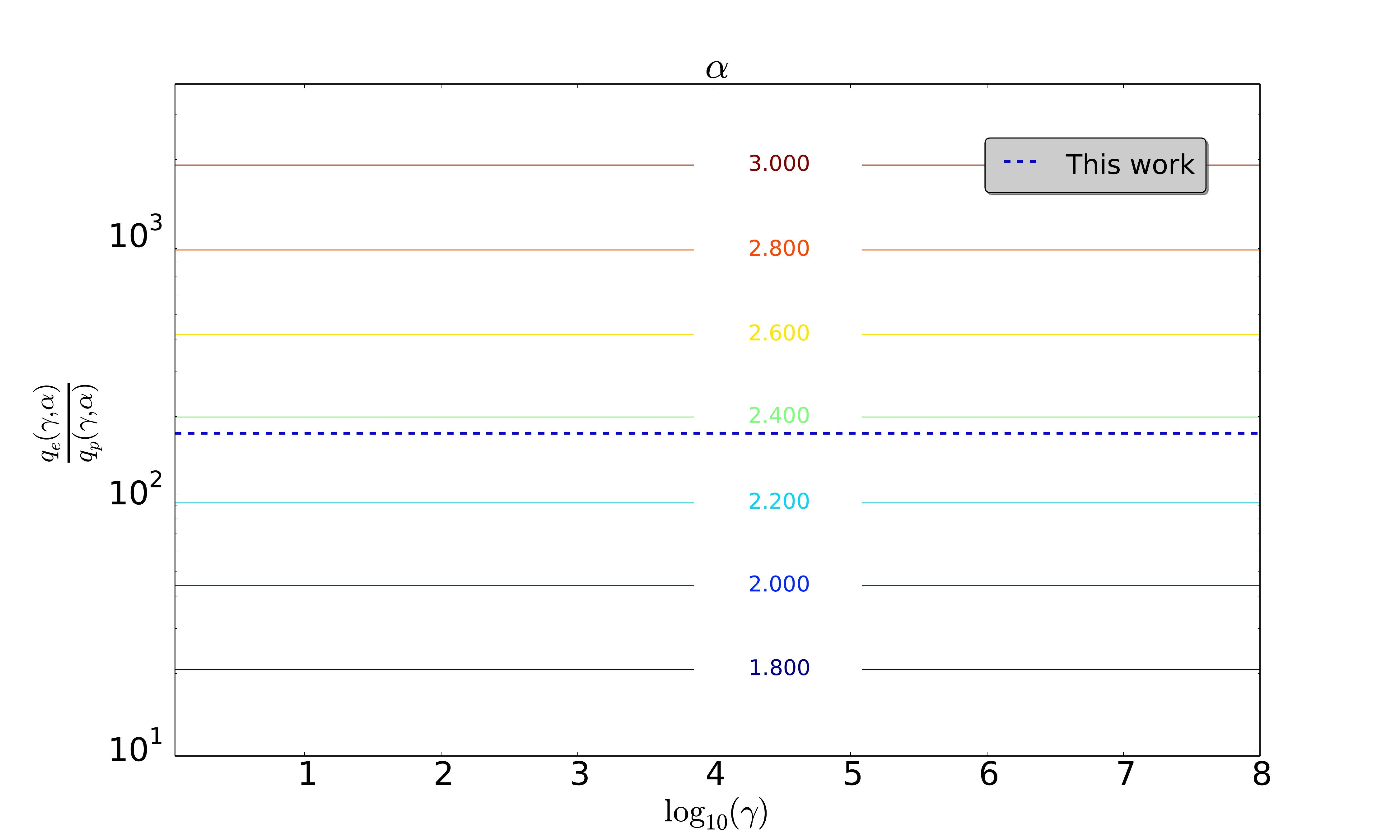}
	\caption[The ratio of primary electrons and protons as a function of the spectral index]{ The ratio of primary electrons and protons as a function of the spectral index $\alpha$ and Lorentz factor $\gamma$, whereby the dashed line represents the relation in Cygnus X.}
	\label{fig:qratiowithoutsecondary2}
\end{figure}

\noindent If a steady state is considered and a homogeneous distribution of CRs without charge imbalances is assumed, the total amount of injected electrons and protons will be the same. Here, we suppose the total acceleration time $T_a$ for electrons and protons is the same.
The total number of accelerated protons or electrons can then be described by:
\begin{equation}
N_0=\int_{0}^{T_a} \rm d t\int_{\gamma_0^i}^{\infty}\rm d\gamma\,q_i(\gamma)=T_a\int_{\gamma_0^i}^{\infty}\rm d\gamma\,q_i(\gamma)
\label{totalNumber}
\end{equation}
for $i= e_1, p$, where $\gamma_0^i$ represents the minimum Lorentz factor and $e_1$ the primary electrons.
Here, it is necessary to distinguish between primary and secondary electrons, as electrons resulting from injection and hadronic interaction are present.\\
Primary electrons $e_1$ denote electrons which received their energy from an accelerator and in the present model obey a power-law, i.e.\ electrons from the injection.\\
In contrast, secondary electrons $e_2$ denote electrons from the decay of muons from hadronic pion production.
\noindent If an effective particle acceleration from shock waves with the velocity $v_s$ is assumed, the particles can be accelerated when their kinetic energy is at least $E_{min,kin}=4\, (\frac{m_p\,v_s}{2})$ which equals 10 keV considering a velocity of $ v_s= 700$ km/s (\cite{Bell}). Given this energy gain, the minimum Lorentz factor becomes:
\begin{equation}
\gamma_0^i=1+ \frac{10 \text{ keV}}{m_ic^2}\ .
\end{equation}
By assuming a power-law spectrum for the injected relativistic particles the related source rate can be expressed by:
\begin{equation*}
q_i(\gamma)=q_0^i\, \gamma(\gamma-1)^{-\frac{\alpha+1}{2}}\ .
\end{equation*}
Considering that the particle energy has an upper limit the source rate becomes:
\begin{equation}
q_i(\gamma)=q_0^i\, \gamma(\gamma-1)^{-\frac{\alpha+1}{2}}\,H[\gamma_{max}-\gamma]\,H[\gamma-\gamma_{min}],
\label{SourceRate}
\end{equation}
where $q_0^i$ denotes the source rate normalization factor and $\alpha$ the energy spectral index\footnote{Notice that the source rate normalization factor in the momentum space as in the work of \cite{Pohl1}, differs by a factor of $m_i^{-1}$.}.%, which is in the supposed to be around 2.7 for galactic extended sources.??
\\Since in Cygnus X many accelerators reside which may complement each other, it is of use to relate the maximal Lorentz factor to the magnetic field and adapt it to the observed spectrum. Hence, $\gamma_{max}^p=10^{13}\times(\frac{B}{G})$ and $\gamma_{max}^e=(\frac{m_p}{m_e})\times\gamma_{max}^p$ is obtained.\\
With eq.(\ref{totalNumber}) and eq.(\ref{SourceRate}), a relation between the electron and proton source rate due to the normalization factor can simply be established.
\begin{equation}
\eta(\alpha)=\frac{q_0^e}{q_0^p}=\frac{\left( (\gamma_{max}^p)^2-1\right) ^{\frac{1+\alpha}{2}}-\left( (\gamma_{0}^p)^2-1\right) ^{\frac{1+\alpha}{2}}}{\left( (\gamma_{max}^e)^2-1\right) ^{\frac{1+\alpha}{2}}-\left( (\gamma_{0}^e)^2-1\right) ^{\frac{1+\alpha}{2}}}\, ,
\label{qRatio}
\end{equation}
which leads for very high energies and an unbroken power-law to a constant ratio of
\begin{equation}
\frac{q_0^e}{q_0^p}\simeq (\frac{m_p}{m_e})^{\frac{\alpha-1}{2}}.
\end{equation}
As a function of the Lorentz factor the electron-proton source rate ratio would lead to
\begin{equation}
\frac{q_p(\gamma)}{q_e(\gamma)}\simeq (\frac{m_p}{m_e})^{\frac{\alpha+1}{2}}
\end{equation} 
or in the space of the momentum $q_i(\gamma)\rightarrow q_i(p)$ which due to the fact $p_i=\gamma\, m_i$, leads to (\cite{Schlicki,Pohl1})
\begin{equation}
\frac{q_p(p)}{q_e(p)}\simeq (\frac{m_p}{m_e})^{\frac{\alpha-1}{2}}\, ,
\end{equation} 
where $p$ denotes the momentum.\\
The ratio of the electron and proton source rates as a function of the spectral index and Lorentz factor $\gamma$  is shown in figure \ref{fig:qratiowithoutsecondary2}. Here, $q_e(\gamma)$ denotes only the primary electron source rate.
\noindent As it will be shown later, the total number of electrons in Cygnus X is composed predominantly of primary electrons. Since this is the case, figure \ref{fig:qratiowithoutsecondary2} also exhibits the ratio of the source rate of all electrons and protons. The quotient of the normalization factor in Cygnus X leads to the consequence $q_p(\gamma)=172\cdot q_ e(\gamma)$, i.e.  the total injection rate of protons is 172 times greater than the injection rate of electrons.\\%If electrons and protons are not injected with the same spectral index, or if they are injected with different minimal energies, the value for $q_e/q_p$ may change significantly (Lukas). As there is no concrete evidence for differences in the spectrum the standard number is used.
\section{Solution of the transport equation}
\noindent The specification of the general solution of electrons or protons relies on the adjustment of the loss processes. The catastrophic losses are already specified and distinguished between electrons and protons. The remaining loss mechanism is the continuous loss, which likewise makes a distinction between protons and electrons since they are based on different interactions and the particles have different rigidities.
\subsection{Solution for electrons}
\noindent Electrons are subject to many losses, which are individually distinctive in different energy ranges. All of the following loss processes will be considered (\cite{Bjoern}):
\begin{flalign}
&\bullet\Lambda_{ion}^e(N_t)\simeq 7.2\cdot 10^{-13} \left( \frac{N_t}{\text{cm}^{-3}}\right) \, s^{-1}\, ,\\
&\bullet\Lambda_{syn}(B)\simeq 1.3\cdot10^{-9}\left( \frac{B}{\text{Gauss}}\right)^2\, s^{-1}\, ,\\
&\bullet\Lambda_{Br}(N_t)\simeq 10^{-15}\left( \frac{N_t}{\text{cm}^{-3}}\right)\, \text{s}^{-1}\, ,\\
&\bullet\Lambda_{IC}(U_{IR})\simeq 5.2\cdot10^{-29}\left( \frac{U_{IR}}{\text{erg}\, \text{cm}^{-3}}\right)\,\left( \frac{R^3}{cm}\right)\, \text{s}^{-1}\, .
\end{flalign}
Here, $ion$ denotes the ionization loss, $syn$ synchrotron loss, $Br$ non-thermal Bremsstrahlung and $IC$ Inverse Compton loss.
In this context $N_t$ represents the constant target density in the plasma, $B$ the predominant magnetic field, $R$ the radius and $U_{IR}$ the infrared energy density.
Hence, the total continuous loss for electrons yields:
\begin{equation}
\Gamma_e \simeq \left( \Lambda_{IC}(U_{IR})+\Lambda_{syn}(B)\right) \gamma^2  +\Lambda_{Br}(N_t)\,\gamma\, + \, \Lambda_{ion}^e(N_t)\ .
\label{ContLoss}
\end{equation}
The associated progression can be found in section \ref{CCLosses}.
Using the variation of constant and skilled integration, we obtain the following expression for the differential CR particle density (\cite{Bjoern}):
\begin{equation}
\begin{split}
n_e(\gamma)=\frac{\Lambda_{ion}^e\exp\left(\chi_e(\gamma)+T_{diff}^e(\gamma) \right)}{(\Lambda_{IC}+\Lambda_{syn})\gamma^2+\Lambda_{Br}\gamma+\Lambda_{ion}^e}  \\
\cdot\int_{\gamma_l}^{\gamma_{max}^e} \rm d\gamma'\, \frac{q_0^e\, \gamma'(\gamma'-1)^{\frac{\alpha-1}{2}}}{(\Lambda_{IC}+\Lambda_{syn})\gamma'^2+\Lambda_{Br}\gamma'+\Lambda_{ion}^e}\\ \cdot\exp\left(-(\chi_e(\gamma')+T_{diff}^e(\gamma' )\right)\, ,
\end{split}
\end{equation}
\begin{equation}
\chi_e(\gamma)=\int\frac{\rm d\gamma}{\Gamma_e\,\tau_{adv}}=\frac{2}{\sqrt{4 a c-b^2 }} \arctan\left( \frac{b + 2 a \gamma}{\sqrt{4 a c-b^2} }\right) 
\label{Advection}
\end{equation}
and
\begin{align}
\begin{split}
T_{diff}^e=&\frac{c\;\lambda_e}{R^2}\int \rm d\gamma\frac{\gamma^{\beta}}{a\gamma^2+b\gamma+c} 
\\=&\frac{c\;\lambda_e}{R^2}\cdot \frac{a}{2\beta\theta} \cdot \big[ (2a\gamma+b-\theta)^{\beta}\cdot\:F_1\left(-\beta,-\beta;1-\beta;\: \frac{b-\theta}{2a\gamma+b-\theta}\right)\\& -(2a\gamma+b+\theta)^{\beta}  \cdot F_1\left( -\beta,-\beta;1-\beta;\: \frac{b+\theta}{2a\gamma+b+\theta}\right) \big]\, .
\end{split}
\label{Diffusion}
\end{align}
Here, the function $F_1$ represents the hyper-geometric function with
\begin{equation}
F_1(-\beta,-\beta,\, 1-\beta, z)=\sum_{k=0}^{\infty}\frac{z^k}{k!}\prod_{k=0}^{\infty}\frac{(-1)(\beta+k)^2}{\beta-1+k}%=e^z\sum_{k=0}^{\infty}\prod_{k=0}^{\infty}\frac{(-1)(\beta+k)^2}{\beta-1+k}
\end{equation}
For the lower integration limit, two cases must be considered
\begin{align}
\gamma_l =  
\begin{cases}
\  \gamma\, , \ \ \ \ \ \ \ \ \text{for}\ \gamma_{min}^e<\gamma<\gamma_{max}^e\, ,\\
\ \gamma_{min}^e\, , \ \ \ \ \text{for}\ \gamma<\gamma_{min}^e\, .
\end{cases}
\label{gamma_l}
\end{align}
In the following the minimum Lorentz factor is supposed to be $\gamma_{min}=1$ and the maximal Lorentz factor the same as supposed in Section \ref{RelationEP} above.
\subsection{Solution for protons}
\noindent In contrast to electrons, protons are also subject to strong interaction. Hence, after an inelastic collision, they can produce a meson by generating a quark anti-quark or change the flavor due to the weak interaction. So, protons are influenced by different losses than electrons, which are also individually distinctive in various energy ranges.\\
In this vein, the ionization loss for the protons as well as the hadronic pion production will be considered (e.g. \cite{SchlickiKrakau2015}):
\begin{equation}
\begin{split}
\Gamma_{p,\pi}&\simeq4.4\cdot 10^{-16}\cdot\left( \frac{N_t}{cm^{-3}}\right) \gamma^{1.28}(\gamma+187.6)^{-0.2}\: s^{-1}\\ &= \Lambda_{p,\pi}(N_t)\cdot\gamma^{1.28}(\gamma+187.6)^{-0.2}\, .
\end{split}
\end{equation}
The proton ionization loss is given by (e.g. \cite{Bjoern}):
\begin{equation}
\Lambda^p_{ion}(N_t)\simeq 1.9\cdot 10^{-16}\cdot\left( \frac{N_t}{cm^{-3}}\right) \: s^{-1} \ .
\end{equation}
Hence, the total loss rate is obtained as the sum of the two:
\begin{equation}
\Gamma_p\simeq \Lambda_{p,\pi}\cdot\gamma^{1.28}(\gamma+187.6)^{-0.2}+\Lambda^p_{ion}.
\end{equation}
The associated progression can be found in section \ref{CCLosses}.
Considering the same procedure as for electrons the differential CR particle density yields (\cite{Bjoern}):
\begin{equation}
\begin{split}
n_p(\gamma)=&\frac{\exp\left( T_{diff}^p(\gamma)+\chi_p(\gamma)\right)}{\Lambda_{p,\pi}\cdot\gamma^{1.28}(\gamma+187.6)^{-0.2}+\Lambda_{io}} \\
&\cdot \int_{\gamma_l}^{\gamma^p_{max}} \rm d\gamma' q_0^p\, \gamma'(\gamma'-1)^{\frac{\alpha-1}{2}}\\&\cdot\exp\left( -T_{diff}^p(\gamma)-\chi_p(\gamma)\right)\, .
\end{split}
\end{equation}
\begin{equation}
\begin{split}
\chi_p(\gamma)=&\int \frac{\rm d\gamma}{|\Gamma|_p\,\tau_{adv}}\\
\simeq &\frac{1}{\Lambda_{p,\pi}\tau_{adv}} \frac{3.571(\gamma+187.6)^{-0.2}}{(0.00533\gamma+1)^{0.2}\gamma^{0.28}}\\
&\cdot F_1\left( -0.28,\ -0.2;\ 0.72;\ -0.00533\, \gamma\right) 
\end{split}
\end{equation} 
and
\begin{equation}
\begin{split}
&T_{diff}^p\simeq\frac{c\lambda_p}{R^2\,\Lambda_{p,\pi}}\frac{(\gamma+187.6)^{-0.2}\gamma^{\beta-0.28}}{(\beta-0.28)(0.00533\gamma+1)^{0.2}}\\
&\cdot F_1\left( -0.2,\ \beta-0.28;\ 1+(\beta-0.28);\ 0.00533\gamma\right)\, .
\end{split}
\end{equation}
The lower integration limit will be considered  in the same way as before.
\section{Radiation processes}
\noindent To understand the radiation from the Cygnus region, it is essential to derive the theoretical expressions for the most critical processes in the astrophysical context.
In the following, an expression for the theoretical flux of each is given.
%\begin{tcolorbox}[breakable, enhanced,colback=black!5!white,colframe=black!40!white,title=\textbf{Intermezzo}]
The theoretical differential flux $\Phi(\gamma)$ can be described as a function of the emissivity and source function  $\varepsilon(\gamma)$, respectively, by (\cite{RadiationProcesses}) :
\begin{equation}
\Phi_{i}(E_j)=\frac{V}{4\pi\,d^2}\varepsilon_i(E_j)\ .
\label{differentialFLux}
\end{equation}	
Here, $i= IC, \, Br,\, \pi^0$ represents an individual process equal Inverse Compton, non-thermal Bremsstrahlung or hadronic $\pi^0$ decay and $j=\gamma, \nu, e^+ ,e^-$ the radiation type. The factor $(4\pi\,d^2)^{-1}$ is a correction for a fraction of the emission, which reaches the observer. The total $\gamma$-ray differential flux is obtained by summation over all present processes.
\begin{equation}
\Phi_{\gamma}(E_{\gamma})=\frac{V}{4\pi d^2}\left( \varepsilon_{IC}(E_{\gamma})+\varepsilon_{Br}(E_{\gamma})+\varepsilon_{\pi^0}(E_{\gamma})\right)\ .
\end{equation}
In addition, the integral flux is given by:
\begin{equation}
\phi_i(E)=\int_{E}^{\infty}\rm d E' \Phi_{i}(\gamma')\ .
\end{equation}
The source function in cm$^{-3}$s$^{-1}$ eV$^{-1}$ must be found individually for each process. It has the following proportionality:
\begin{equation}
\varepsilon_{p_1,\,p_2}\propto c\int_{\gamma_{min}}^{\infty}\rm d\gamma\,n_{p_1}(\gamma)\,n_{p_2}(\gamma)\frac{d\sigma_{p_1,\,p_2}}{d\gamma}\, .
\label{emissivity}
\end{equation}
It mainly depends on the differential density $n_i(\gamma)$ of the interacting particles $i=p_1,\ p_2$ and the related differential cross section $\rm d\sigma_{p_1,\,p_2}/{\rm d\gamma}$.
%	\end{tcolorbox}
%	\bigskip
\subsection{Synchrotron radiation}
\noindent To use only synchrotron radiation as the vital radio emission process and to avoid free-free emission and Bremsstrahlung, respectively, only the non-thermal emission from Cygnus X will be taken into account by considering a radio spectrum $\lesssim 10\ GHz$ (\cite{Bjoern}).
\noindent The emissivity of synchrotron radiation is given by (\cite{Bjoern, RadiationProcesses}):
\begin{equation}
\varepsilon_{syn}(\nu)=\frac{1}{4\pi}\int_{\gamma_{min}}^{\gamma_{max}}n_e(\gamma)P_{syn}(\nu,\gamma)\, d\gamma
\end{equation}
\begin{equation}
\begin{split}
&\text{with}\ \ P_{syn}=P_0\cdot\left( \frac{\nu}{\gamma^2\nu_s}\right)^{1/3}\exp(- \frac{\nu}{\gamma^2\nu_s}) \\ &\text{and}\ \ P_0= 2.65\cdot10^{-10}\cdot\left( \frac{B}{1\, G}\right)\ \text{eV s$^{-1}$ Hz$^{-1}$}.
\end{split}
\end{equation}
In contrast to electrons, protons do not emit synchrotron radiation at the same intensity level, because the emitted power of synchrotron radiation $\dot{E}_{syn}$ is proportional to $ m^{-4}$.
\begin{equation}
\frac{\dot{E}_{syn}^e}{\dot{E}_{syn}^p}= \left( \frac{m_p}{m_e}\right)^4\simeq 1.13\cdot10^{13}
\end{equation}
Since the ratio of the proton-electron mass is \newline $\sim 1836$, the proton synchrotron radiation requires inconceivably high energies and a strong magnetic field (\cite{Proton_Syn}).
\subsection{Inverse Compton}
\noindent In astrophysical context, the Inverse Compton process is based on the interaction of a relativistic CR electron with an ambient photon. The relativistic electron transfers a part of its kinetic energy to the target photon, whereby a minimum energy of 
\begin{equation}
E_{min}= \frac{E_{\gamma,f}}{2}\left[1+\left( 1+\frac{m_e^2c^4}{E_{\gamma,i} E_{\gamma,f}}\right)  \right] 
\end{equation} 
is necessary.  Here, $E_{\gamma,i}$ and $E_{\gamma,f}$ represent the initial and final photon energy respective to the scattering (\cite{Schlicki}). \\ \\
Since Cygnus X contains a large amount of dust, the primary photon field is represented by infrared emission, as the starlight is absorbed and then re-radiated in the infrared range. Thus, $\gamma E_{\gamma,i}\ll m_e\,c^2$ is valid, and the Inverse Compton loss can be considered in the Thomson limit. Only a small fraction of the injected electrons are within the condition $\gamma>m_e\,c^2/E_{\gamma,i}$, the Klein-Nishina (KN) regime. The following further considers a maximal Lorentz factor (see Section \ref{RelationEP} for more details).
Considering a gray body or rather a modified blackbody according to (\cite{Casey}) and an isotropic and uniform spatial distribution, the differential infrared photon density yields (\cite{Bjoern, Casey}):

\begin{equation}
\begin{split}
\frac{\rm{d} \it n_{IR} (E)}{\rm{d} \it E}=&1.125\cdot 10^{19}\cdot\frac{U_{IR}}{E_0}\left( \frac{E}{h\, c}\right) ^3\\&\cdot\frac{1-\exp\left( ({E}/{E_0})^{\beta}\right) }{\exp\left( E/(k_B\,T_D)\right) -1}\ .
\end{split}
\end{equation}
Here, $T_D$ denotes the dust temperature, $U_{IR}$ the infrared photon energy density, $\beta=1.5$ the emissivity index (\cite{Casey}) and $E_0=12.4\cdot10^{-3}$ eV the energy (\cite{Casey}), where the optical depth equals unity. In Cygnus X the dust temperature is supposed to be approximately 25 K (\cite{DustTemp}) and the infrared photon energy density 5 eV/cm$^3$ (\cite{TovaPaper}).
The photon density is then given by
\begin{equation}
\begin{split}
n_{IR}(E)&=1.125\cdot 10^{19}\frac{U_{IR}}{E_0(h\, c)^3}\\&\cdot\int \rm d \it E\: E^3 \frac{1-\exp\left(- ({E}/{E_0})^{\beta}\right) }{\exp(E/k_B\,T_D)-1}.
\end{split}
\end{equation}
According to eq.(\ref{emissivity}) the differential cross section for the Inverse Compton process is needed, which is given by the Klein-Nishina formula:
\begin{equation}
\begin{split}
&\frac{\rm d\sigma(E_{\gamma,f},E_{\gamma,i},\gamma)}{\rm \gamma}=\frac{3}{4}\frac{\sigma_T}{E_{\gamma,i}\gamma^2}G(q,\Gamma), \ \text{with}\\
&G(q,\Gamma)= 2q\ln(q)+(1-q)(1+2q)+\frac{(\Gamma q)^2(1-q)}{2(1+\Gamma q)}\\
&q=\frac{E_{\gamma,f}}{\Gamma(\gamma m_e\,c^2-e_{\gamma})}\ \ \, \Gamma =\frac{4\gamma E_{\gamma,i}}{m_e\,c^2}\ ,
\end{split}
\end{equation}
\noindent whereby $\gamma$ denotes the electron Lorentz factor. According to eq.(\ref{emissivity}) the emissivity yields
\begin{equation}
\varepsilon_{e,\,\gamma}^{IC}(E_{\gamma})=\frac{3}{4}c\,\sigma_T\,n_{IR}\int_{\frac{E_{min}}{m_ec^2}}^{\infty} \frac{\rm d\gamma}{\gamma^2} n_e(\gamma)G(q,\Gamma)\ .
\end{equation}
\subsection{Non-thermal Bremsstrahlung}
\noindent In an astrophysical context, Bremsstrahlung is produced in a hot and predominantly ionized plasma, where the particles are free before and after the deflection. Bremsstrahlung is an important process in Cygnus X, since it contains H$_{\text{II}}$ regions, ionized gas clouds around hot and young stars (\cite{IdentificationTeVCygnusCocoon,HighEnergyAstrophysics}).\\
The general emissivity produced by Bremsstrahlung can be described by (\cite{Stecker}):
\begin{equation}
\varepsilon^{Br}_{e,\,\gamma}(E_{\gamma})=\int_{\gamma}^{\infty}d\gamma\,N_t\,c\,\frac{\sigma_{Br}}{E_{\gamma}}n_e(\gamma)=\frac{N_t\,c\,\sigma_{Br}}{E_{\gamma}}\int_{E_{\gamma}/m_e}^{\infty}d\gamma\,n_e(\gamma)\ .
\end{equation}
Here, $\sigma_{Br}=3.38\cdot 10^{-26}$ cm$^2$ denotes the Bremsstrahlung cross section and $N_t$ the proton target density. The flux can be calculated using eq.(\ref{differentialFLux}). 
\subsection{Gamma-rays from hadronic pion production }
\noindent After the interaction of protons with the ambient medium pions are generated. Thus, the inelastic proton-proton cross section $\sigma_{pp, inel}$ must be considered by equation(\ref{ppCrossSection}) (\cite{Kelner}).
In succession, the $\pi^0$ decays and generates two $\gamma$-rays. This process is thought to cause most of the $\gamma$-rays in star forming regions (\cite{Bjoern}).\\
Additionally, the energy spectrum $F_{\gamma}(x,E_p)$ of secondary particles (here $\gamma$-rays), which are produced in one interaction with a proton of the energy $E_p$, must also be considered. This includes the intermediate production and decay of $\pi^0$, whereby $x_{\gamma}=E_{\gamma}/E_{\pi}$ is the ratio of the energy of the incident proton to the produced $\pi^0$. In this regard $F_{\gamma}(x_{\gamma},E_p)\cdot dE_{\gamma}/E_p$ is the number of $\gamma$-rays from a single proton-proton interaction in the interval $[E_{\pi}, E_{\pi}+dE_{\pi}]$ (\cite{Kelner}).
\begin{equation}
\begin{split}
\sigma_{pp, inel}(E_p)=& 
\begin{cases}
(34.3+1.88\,L+0.25\,L^2)\left[ 1-(\frac{E'}{E_p})^4\right]\ \text{mb,   }   \text{for}\ E'\leq E_p\leq0.1\ \text{TeV} \\
(34.3+1.88\,L+0.25\,L^2)\ \text{mb, }  \ \ \ \ \ \qquad\qquad\text{for}\ E_p>0.1\ \text{TeV}\ 
\end{cases}\\
&\text{with}\ E'=\left( m_p+2m_{\pi} +m_{\pi}^2/2m_p\right) c^2 \ \text{  and  }\  L=\ln(E_p/1\ \text{TeV})\, . \ \ \
\end{split}
\label{ppCrossSection}
\end{equation}
Using this relation, the total emissivity is obtained by:
\begin{equation}
\varepsilon_{p,\pi^0,\gamma}^{had}(E_{\gamma})=c\, N_t\int_{E_{\gamma}}^{\infty}\frac{dE_p}{E_p}\sigma_{pp, inel}(E_p) \cdot n_p(E_p)F_{\gamma}(x_{\gamma},E_p)\ .
\end{equation}
\subsection{Neutrinos}
\noindent In the astrophysical context, neutrinos can be produced from hadronic charged pion production and after that from leptonic muon decay.\\
In the same manner, as the emissivity was determined for $\gamma$-rays from the hadronic pion decay by considering its energy spectrum, the energy spectra can be replaced for another particle $i$, i.e.  $F_{\gamma}(E_{\gamma}/E_{\pi^0},\ E_p)\longmapsto F_{i}(E_{i}/E_{\pi^{\pm}},\ E_p)=F_i(x_i,E,p)$
for  $i=\nu_e,\nu_{\mu}, e$.\\
Since $F_{i}(x_{i},E_p){dE_{\pi}}/{E_p}$ denotes the number of the particle $i$ for a single proton-proton interaction in the interval $[E_{\pi}, E_{\pi}+dE_{\pi}]$, the following approximation is valid: $F_e(x_e,\ E_p)\simeq F_{\nu_e}(x_{\nu_e},\ E_p)$. The deviation is less than 5\% (\cite{Kelner}) when using this approximation. 
\noindent If the muon neutrino from the process
\begin{equation}
\pi^{\pm}\longrightarrow \mu^{\pm}+\nu_{\mu}/\bar{\nu}_{\mu}
\label{PionDecay}
\end{equation} 
is denoted as $\nu_{\mu1}$ and the muon from the process \begin{equation}
\mu^{\pm}\longrightarrow e^{\pm}+ \nu_e/\bar{\nu}_e+\bar{\nu}_{\mu}/\nu_{\mu}
\label{MuyonDecay}
\end{equation}
as $\nu_{\mu2}$, the following can be approximated:\\ $F_e(x_e,E_p)\simeq F_{\nu_{\mu2}}(x_{\nu_{\mu2}},\ E_p)$. 
Hence, the emissivities can be described by:
\begin{equation}
\varepsilon_{p,\pi,\nu_{\mu2}}(E_{\nu_{\mu2}})=c\, N_t \cdot\int_{E_{\nu_{\mu2}}}^{\infty}\frac{dE_p}{E_p}\sigma_{pp, inel}(E_p)\,n_p(E_p)F_{e}(x_{\nu_{\mu2}},E_p),
\end{equation}
\begin{equation}
\varepsilon_{p,\pi,\nu_{e}}(E_{\nu_{e}})=c\, N_t \cdot\int_{E_{\nu_{e}}}^{\infty}\frac{dE_p}{E_p}\sigma_{pp, inel}(E_p)\,n_p(E_p)F_{e}(x_{\nu_{e}},E_p)
\end{equation}
and
\begin{equation}
\varepsilon_{p,\pi,\nu_{\mu1}}(E_{\nu_{\mu1}})=c\, N_t\cdot\int_{E_{\nu_{\mu1}}}^{\infty}\frac{dE_p}{E_p}\sigma_{pp, inel}(E_p)\,n_p(E_p)F_{\nu_{\mu1}}(x_{\nu_{\mu1}},E_p)\ .
\end{equation}
The total neutrino emissivity is the sum of the contributions:
\begin{equation}
\varepsilon_{p,\pi,\nu}(E_{\nu})=c\, N_t\int_{E_{\nu}}^{\infty}\frac{dE_p}{E_p}\sigma_{pp, inel}(E_p)\,n_p(E_p)\cdot\left( F_{\nu_{\mu}}(x_{\nu},\, E_p)+2F_{e}(x_{\nu},\, E_p)\right)\ ,
\end{equation}
where we used the differential particle density of relativistic protons from Section \ref{Model}.\\
Hereby, the ratio of the appearance of different type of neutrino at the source is (1:2:0) for \newline($\nu_e:\nu_{\mu}:\nu_{\tau}$). Due to the neutrino oscillation at the large distance the ratio changes to 1:1:1 (e.g. \cite{Julia1}), whereby this condition is fulfilled for Cygnus X.\\
\subsection*{Secondary electrons}
\noindent Analogously to the case of the neutrino, the emissivity for secondary electrons which are generated from hadronic pion production is calculated with
\begin{equation}
\varepsilon_{p,\pi,{e}}(E_{{e}})=c\, N_t\int_{E_{{e}}}^{\infty}\frac{dE_p}{E_p}\cdot\sigma_{pp, inel}(E_p)\,n_p(E_p)F_{e}(x_{{e}},E_p)\, .
\label{sFktElectron}
\end{equation}
\section{Primary and secondary electrons in Cygnus X}
\begin{figure}[H]
	\centering
	\subfigure{\includegraphics[width=0.9\linewidth]{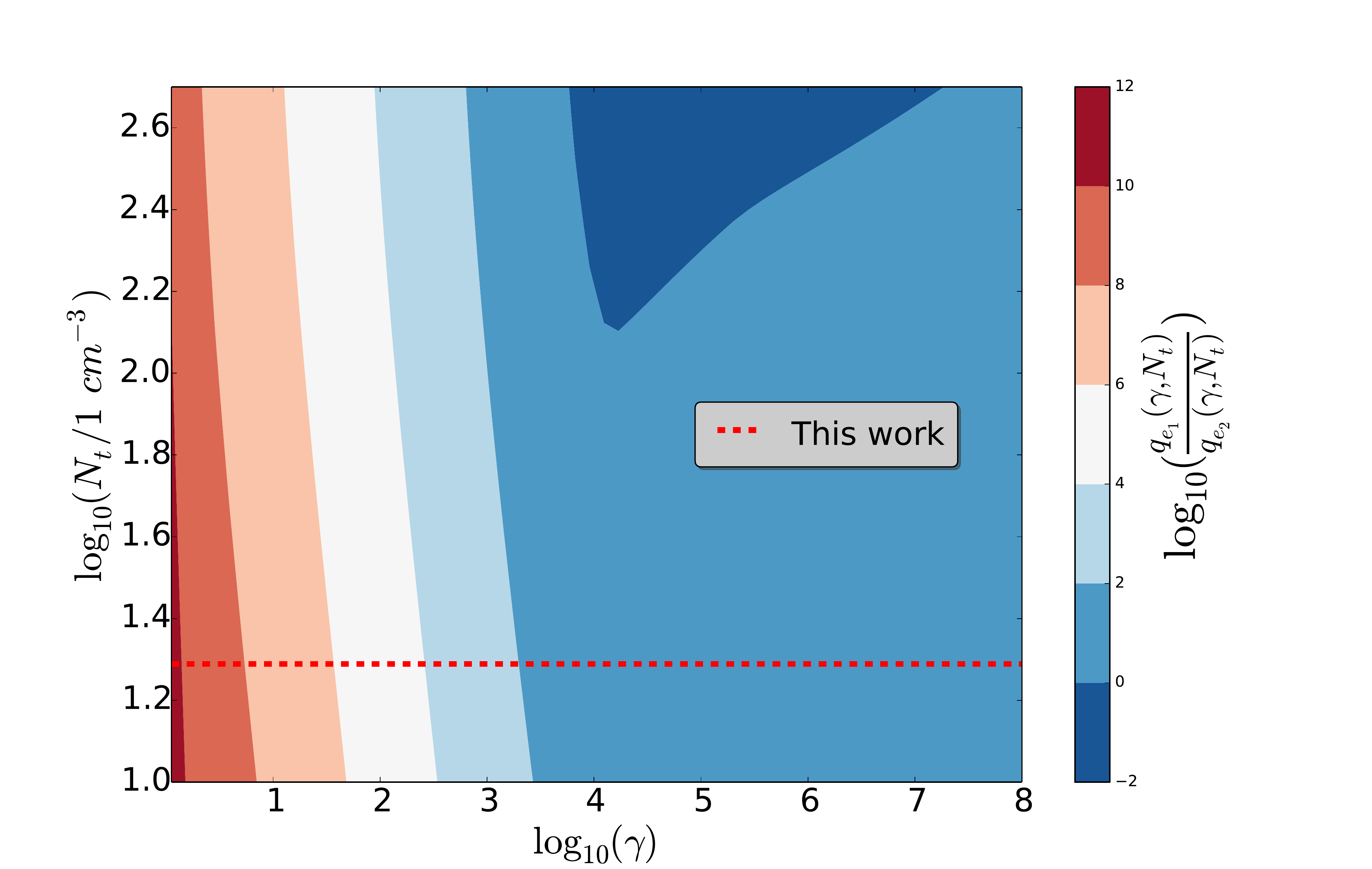}}
	\caption[Ratio of primary and secondary electrons as a function of the target density]{The ratio of primary and secondary electrons $q_{e1}(\gamma,N_t)/q_{e2}(\gamma,N_t)$ as a function of the target density $N_t$ and the Lorentz factor $\gamma$ for $B=8.9\times10^{-6}$ G and $\alpha=2.37$.}
	\label{fig:primarysecondariesNt}
\end{figure}

\begin{figure}[H]
	\centering
	\subfigure{\includegraphics[width=0.9\linewidth]{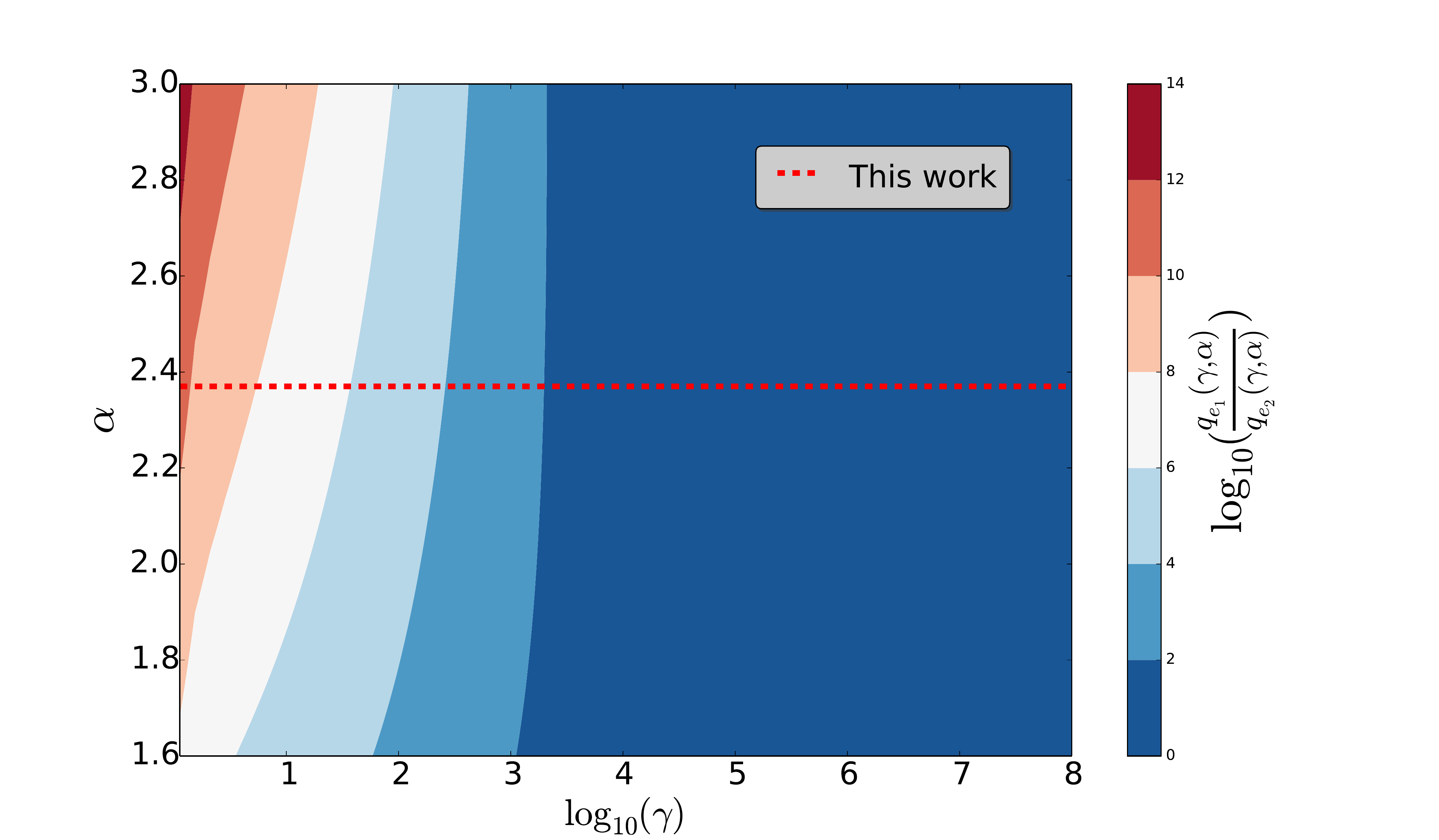}}
	\caption[Ratio of primary and secondary electrons as a function of the spectral index]{The Ratio of primary and secondary electrons $q_{e1}(\gamma,\alpha)/q_{e2}(\gamma,\alpha)$ as function of the spectral index $\alpha$ and the Lorentz factor $\gamma$ for $N_t=19.4$ cm$^{-3}$ and  $B=8.9\times10^{-6}$ G.}
	\label{fig:primarysecondariesalpha}
\end{figure}

\begin{figure}[H]
	\centering
	\subfigure{\includegraphics[width=0.9\linewidth]{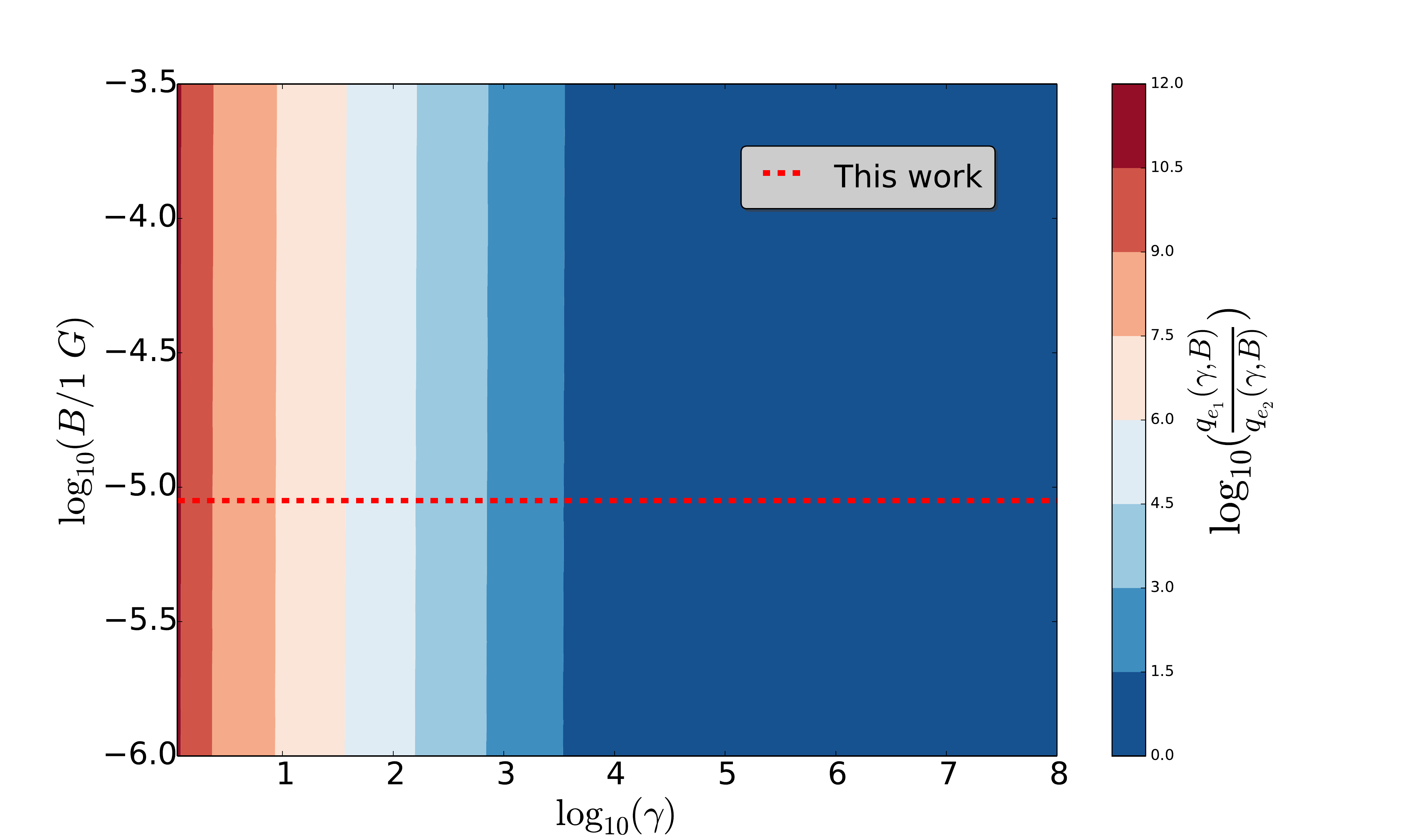}}
	\caption[Ratio of primary and secondary electrons as a function of the magnetic field]{The ratio of primary and secondary electrons $q_{e1}(\gamma,B)/q_{e2}(\gamma,B)$ as a function of the magnetic field $B$ and the Lorentz factor $\gamma$ for $N_t=19.4$ cm$^{-3}$,  $\alpha=2.37$.}
	\label{fig:primarysecondariesB}
\end{figure}

\noindent Because secondary electrons result from proton interaction or rather a hadronic pion production, they depend on the proton density and on its cross section to produce charged pions. Therefore, it also depends on the Lorentz factor and the target density. Since $\gamma_{max}^i$ in our model is related to the magnetic field strength and the primary electron source rate includes the spectral index, the dependence will be plotted to present the ratio of primary to secondary electron source rate as a function of the target density $N_t$, magnetic field strength $B$ and the spectral index $\alpha$. The associated figures can be seen in figures \ref{fig:primarysecondariesNt}-\ref{fig:primarysecondariesB}.
\noindent \noindent Here, $q_{e_1}(\gamma)$ denotes the primary and $q_{e_2}(\gamma)$ the secondary electron source rate function with $q_{e_2}(\gamma)=\varepsilon_{p,\pi,{e}}(E_{{e}})\cdot m_ec^2$. The ratio of primary to secondary electrons as a function of the Lorentz factor for certain parameters which in this work seems to describe Cygnus X, is represented by the dashed red line.\\
The figures \ref{fig:primarysecondariesNt} - \ref{fig:primarysecondariesB} clarify that the amount of secondary electrons increases with greater $\gamma$ relative to the primary electrons. Additionally, the rise for Cygnus X seems to be quite uniform.  In any event, in Cygnus X the primary electrons always dominate. %\newpage
\section{Continuous and catastrophic losses}
\label{CCLosses}
\begin{figure}[H]
	\centering
	\subfigure{\includegraphics[width=0.9\linewidth]{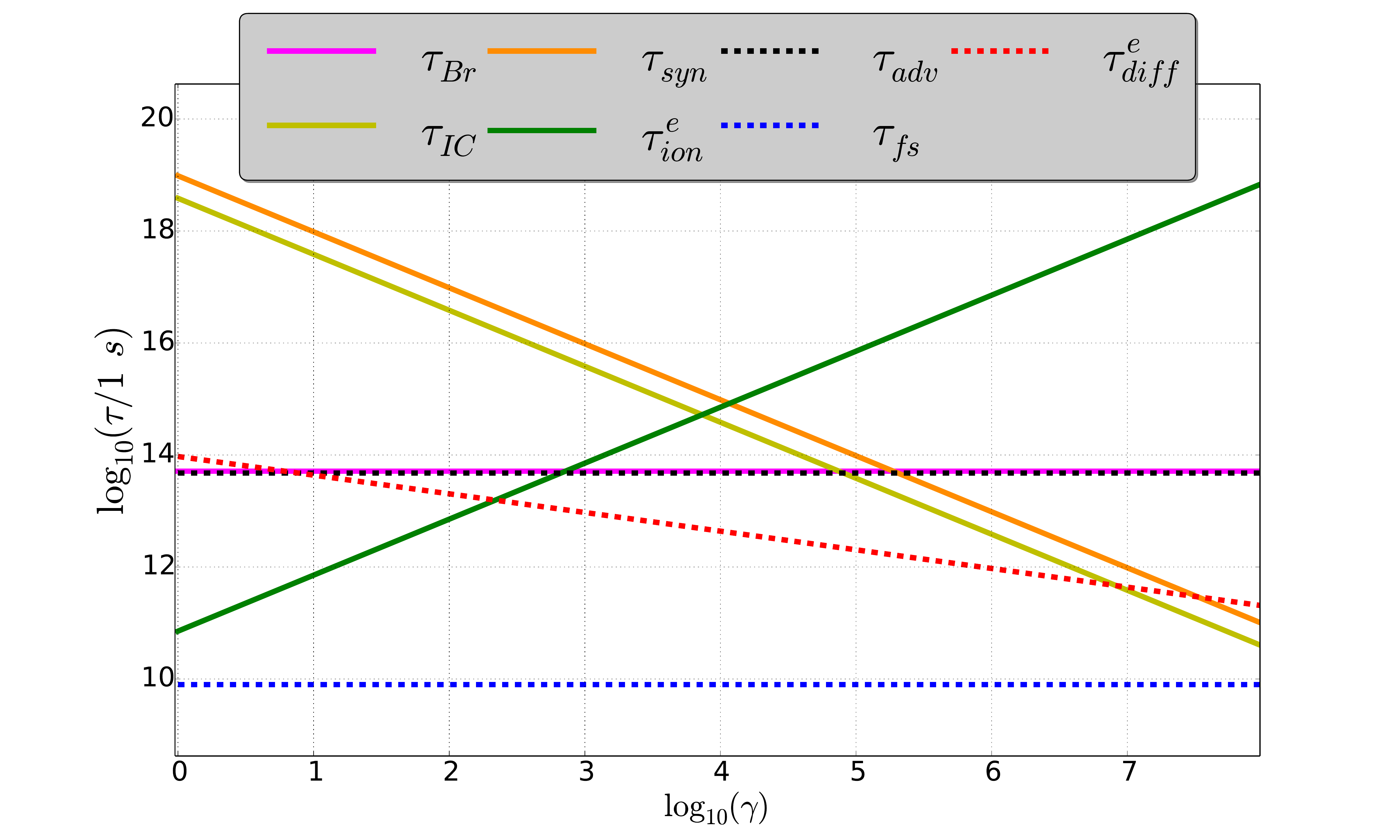}}
	\caption[Continuous and catastrophic timescales for Electrons]{Continuous timescales (solid lines) and catastrophic timescales (dashed lines) for electrons as a function of the Lorentz factor $\gamma$ with $B=8.9\times 10^{-6}$ G, $N_t=19.4$ cm$^{-3}$ and $U_{IR}=$5 eV/cm$^3$.}
	\label{fig:timescaleallP}
	
\end{figure}

\begin{figure}[H]
	\subfigure{\includegraphics[width=0.9\linewidth]{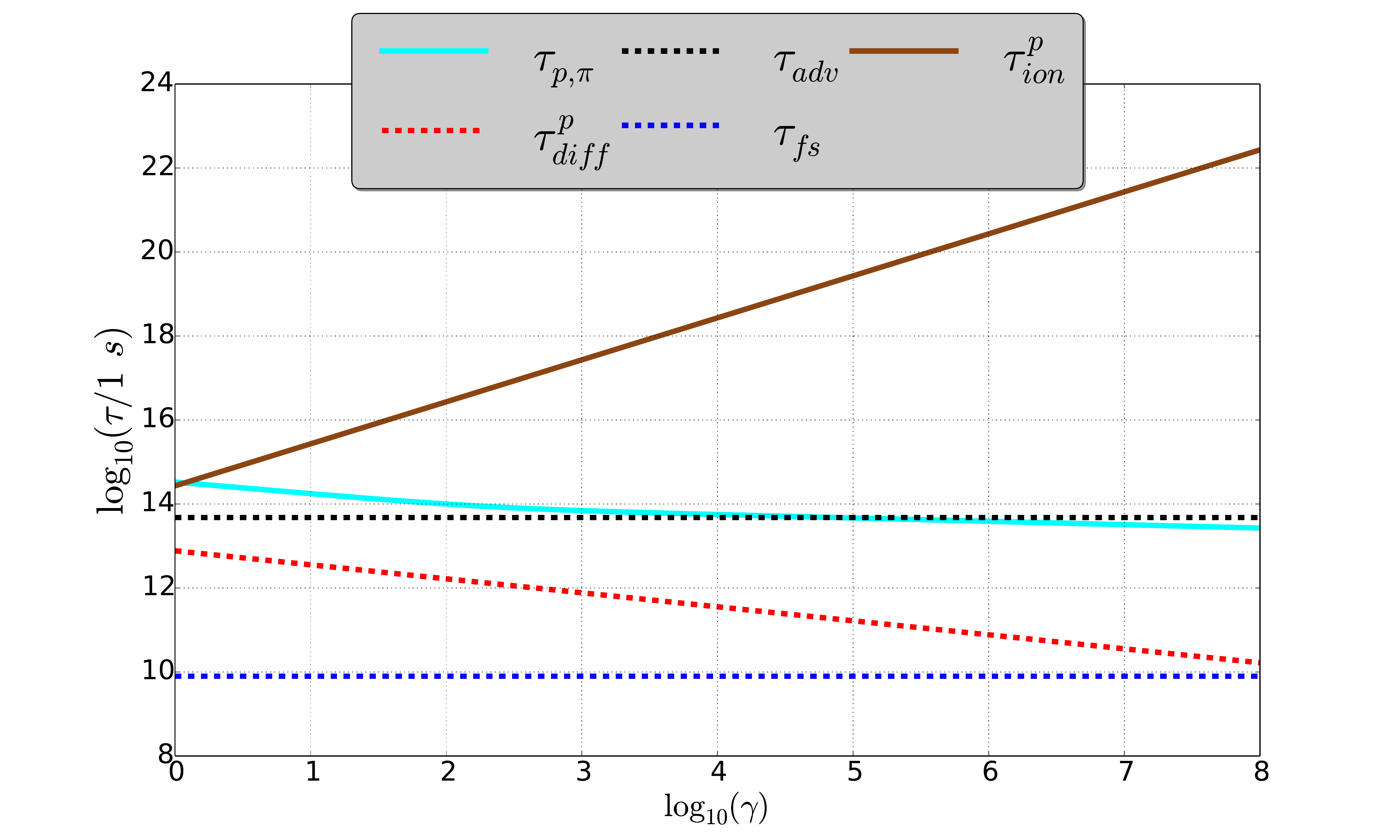}}
	\caption[Continuous and catastrophic timescales for Protons]{Continuous timescales (solid lines) and catastrophic timescales (dashed lines) for protons as a function of the Lorentz factor $\gamma$ with $N_t=19.4$ cm$^{-3}$.}
	\label{fig:timescaleallE}
	
\end{figure}

\noindent Whether the continuous or catastrophic momentum loss adopts the main loss mechanism in Cygnus X or whether both are equally significant can be determined by regarding the related timescale. The total continuous loss timescale is given by:
\begin{equation}
\begin{split}
\varSigma_{con}^e\equiv&\frac{1}{\tau_{con}^e}=\frac{\Gamma_e}{\gamma}=\frac{\Lambda^e_{ion}(N_t)}{\gamma}+ \Lambda_{IC}(U_{IR})\, \gamma\\&+{\Lambda_{syn}(B)\, \gamma} + \Lambda_{Br}(N_t) \\=&\frac{1}{\tau_{ion}^e}+\frac{1}{\tau_{IC}}+\frac{1}{\tau_{syn}}+\frac{1}{\tau_{Br}}
\end{split}
\end{equation}
\begin{equation}
\begin{split}
&\varSigma_{con}^p\equiv\frac{1}{\tau_{con}^p}=\frac{\Gamma_p}{\gamma}=\frac{\Lambda^p_{ion}(N_t)}{\gamma}+\Lambda_{p,\pi}(N_t)\\&\cdot\gamma^{0.28}(\gamma+187.6)^{-0.2}=\frac{1}{\tau_{ion}^p}+\frac{1}{\tau_{p,\pi}}\, .
\end{split}
\end{equation}
%The particles are not allowed to travel faster than the speed of light during their path through Cygnus X. Hence, the following condition must be fulfilled
%\begin{equation}
%\frac{R^2}{c\, \lambda_i\, \gamma^{\beta}}>\frac{R}{c}
%\end{equation}
%If this condition is no longer valid, $\lambda_i\gamma$ will be kept constant.
In figures \ref{fig:timescaleallE} and \ref{fig:timescaleallP} the continuous timescales are represented by solid lines and catastrophic timescales by dashed lines. Here, $\tau_{fs}=R/c$ denotes the timescale of a free streaming particle with the velocity of light c which is needed to pass through the considered region of Cygnus X within 77 pc.\\
As the diffusion timescale represents a quantity for the entrapment of the particle particles in Cygnus X, it must be smaller than $\tau_{fs}$ because the particle cannot move faster than the speed of light. If $\tau_{diff}\approx\tau_{fs}$ then the diffusion is negligible as in this case the particles will move undisturbed.
\noindent For most of the time, the diffusion timescale has the smallest value. In particular, the Inverse Compton and synchrotron timescales are most important at higher energies. They decrease with a similar slope which is according to the amount the strongest. The timescales of Bremsstrahlung and advection do not differ significantly from each other and are constant. \\  
\noindent The importance of diffusion is recognizable for protons as it is always the dominant process. In contrast, ionization loss is even at the lowest energies almost negligible. Moreover, hadronic pion production is the second most important loss mechanism at higher energies and advection is relevant for $\gamma < 10^4$.\\
Thus in total, the diffusion loss timescale for catastrophic losses is smaller than that for advection for all energy ranges. Additionally, the Inverse Compton loss becomes the main loss mechanism for electrons at very high energies.\\

\begin{figure}[H]
	\centering
	\includegraphics[width=1.0\linewidth]{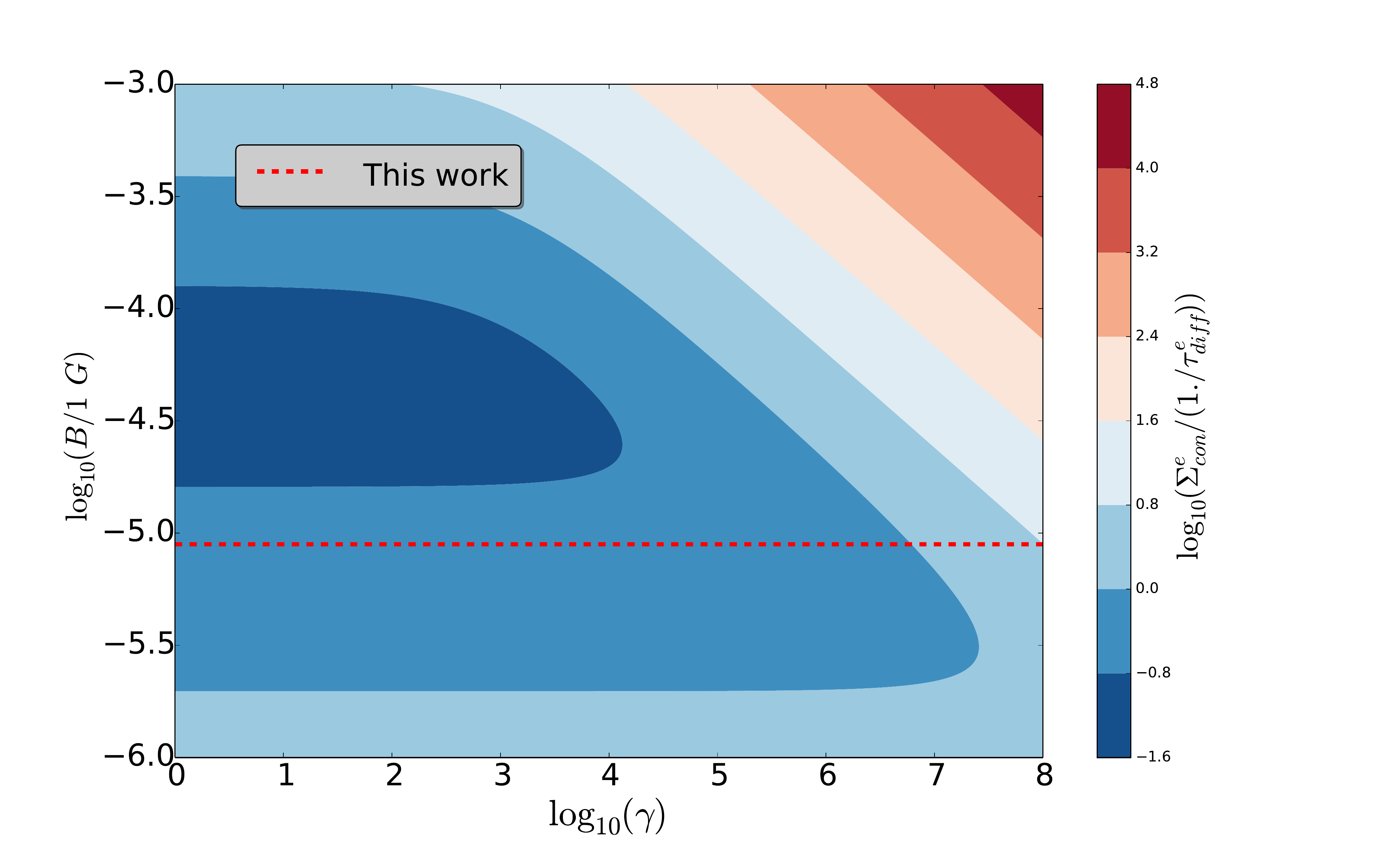}
	\caption[Ratio of continuous and catastrophic losses as a function of the magnetic field for electrons]{The ratio of $\Sigma_{con}^e$ and $1/\tau_{diff}^e$ for electrons as a function of the magnetic field strength $B$ and the Lorentz factor $\gamma$ with $N_t=19.4$; the dashed line shows the dependency in Cygnus X.}
	\label{fig:electronmesh}
\end{figure}
\begin{figure}[H]
	\centering
	\includegraphics[width=0.95\linewidth]{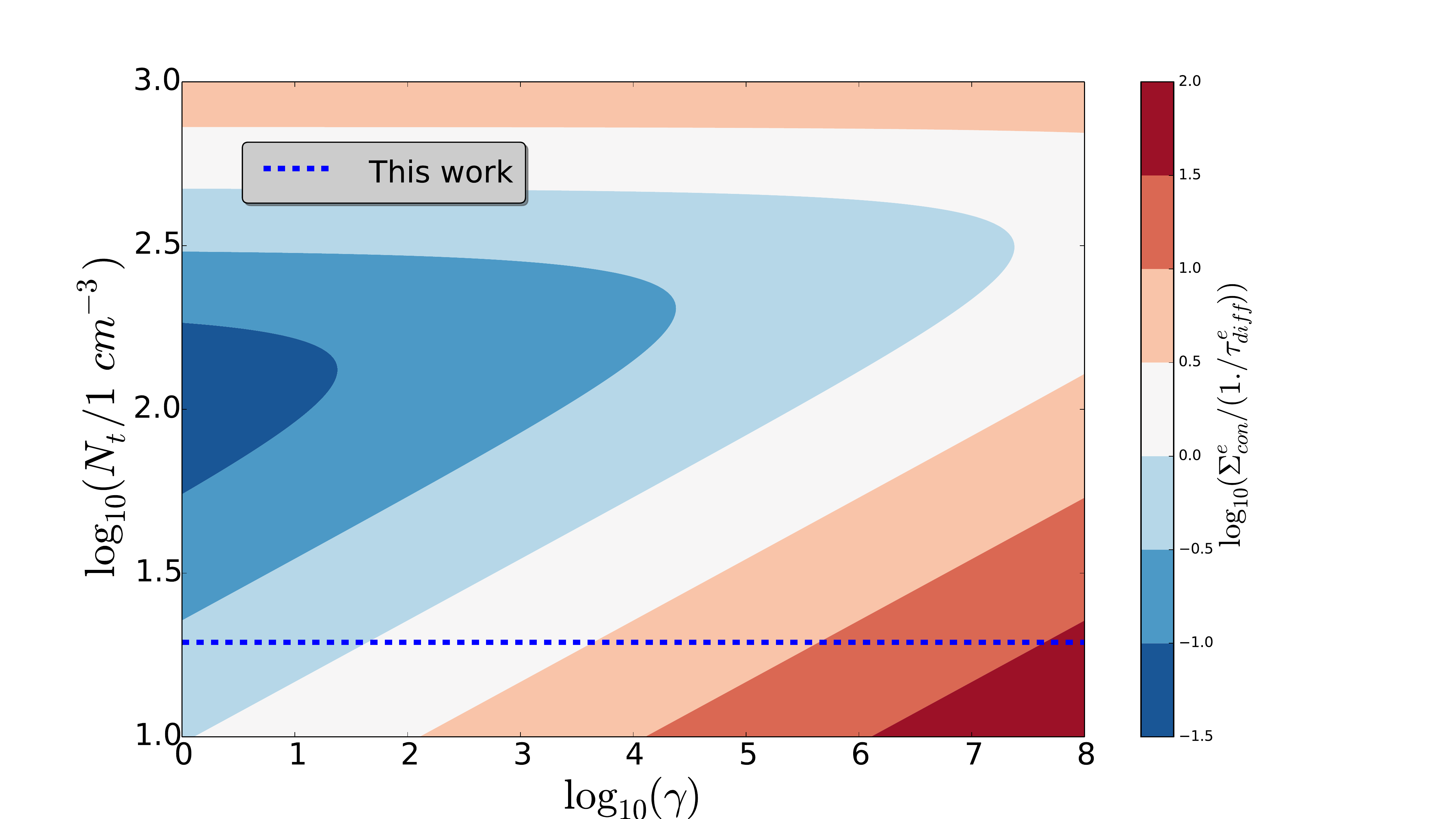}
	\caption[Ratio of continuous and catastrophic losses as a function of the target density for electrons]{The ratio of $\Sigma_{con}^e$ and $1/\tau_{diff}^e$ for electrons as a function of the target density $N_t$ and the Lorentz factor $\gamma$ with $B=8.9\times10^{-6}$ G; the dashed line shows the dependency in Cygnus X.}
	\label{fig:electronmesh1}
\end{figure}
\begin{figure}[H]
	\centering
	\includegraphics[width=0.95\linewidth]{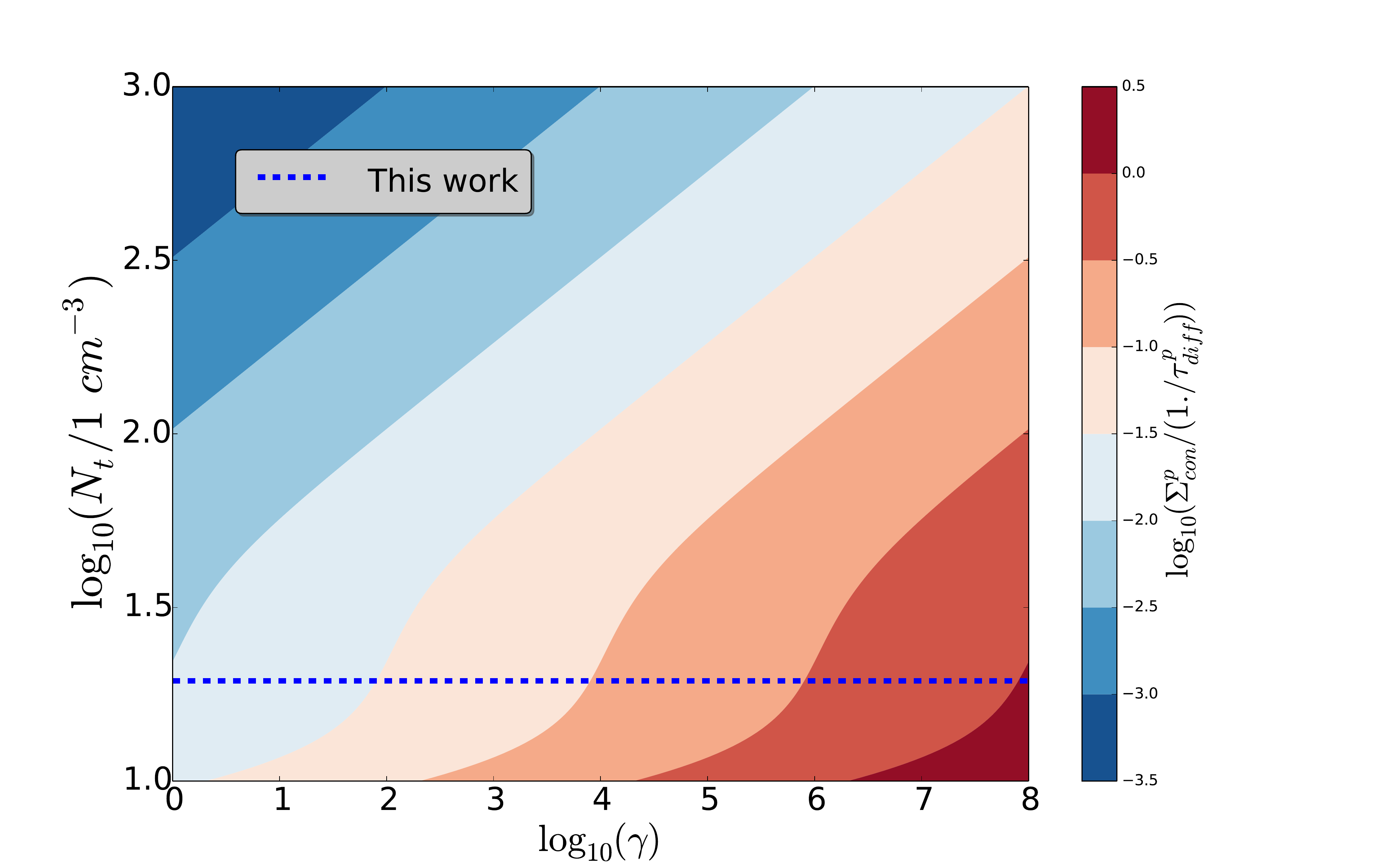}
	\caption[Ratio of continuous and catastrophic losses as a function of the target density for protons]{The ratio of $\Sigma_{con}^p$ and $1/\tau_{diff}^p$ for protons as a function of the target density $N_t$ and the Lorentz factor $\gamma$; the dashed line shows the dependency in Cygnus X.}
	\label{fig:protonmesh}
\end{figure}
\begin{figure}[H]
	\centering
	\subfigure[]{\includegraphics[width=0.8\linewidth]{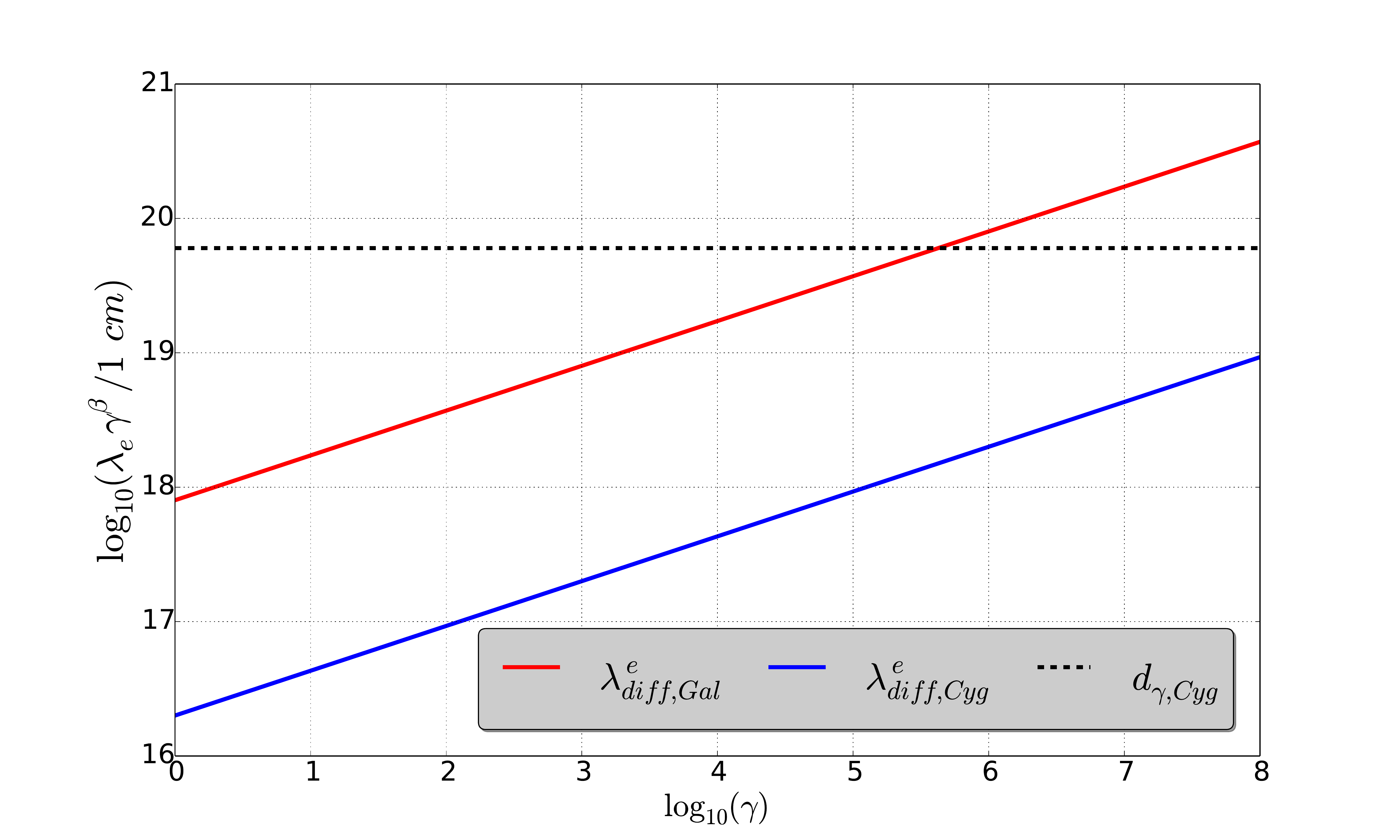}}
	\subfigure[]{\includegraphics[width=0.8\linewidth]{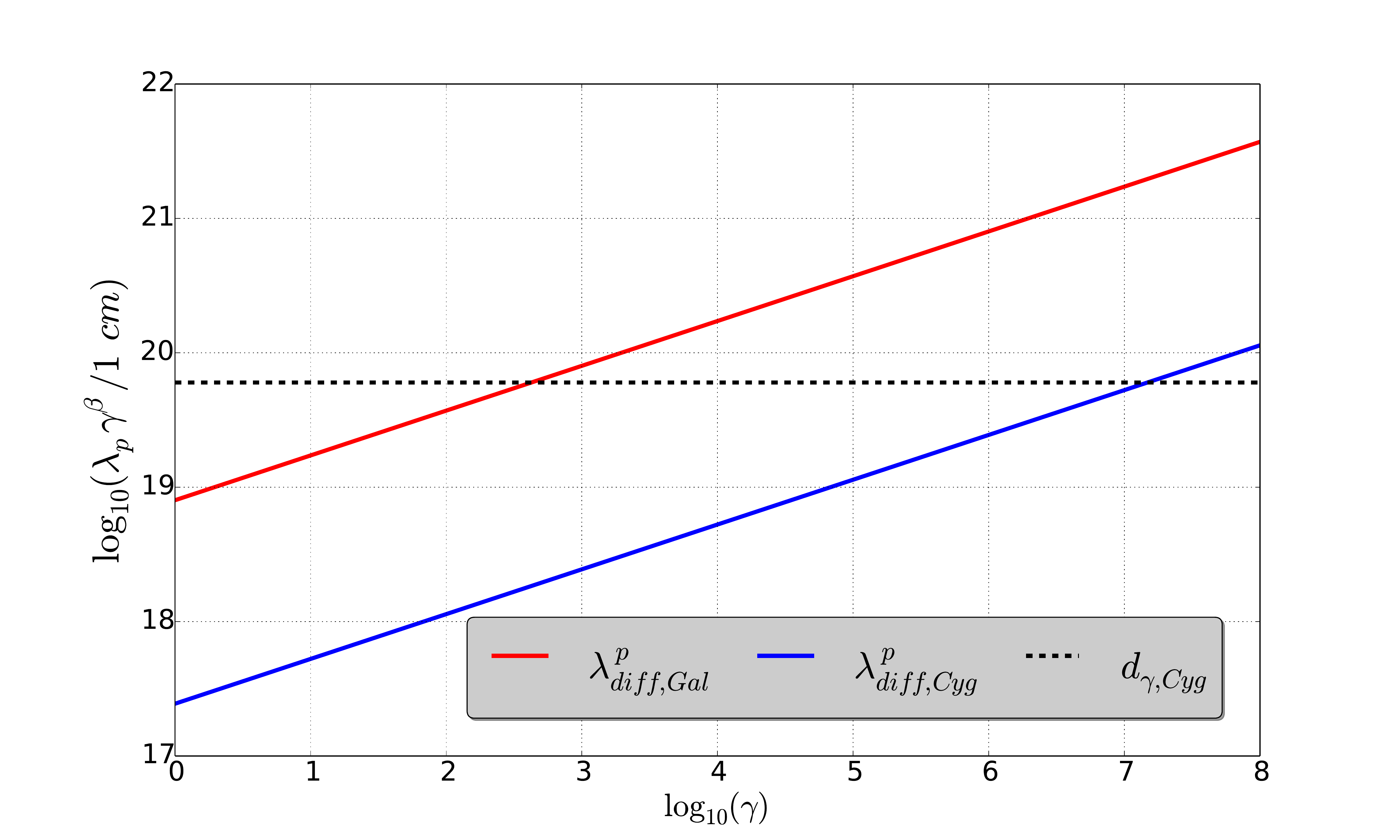}}
	\caption[Mean free path in Cygnus X and in the Galaxy]{Mean free paths in Cygnus X and  the Galaxy for electrons and protons as a function of the Lorentz factor.}
	\label{fig:cyggal}
\end{figure}

\newpage
\noindent It is meaningful to investigate when the continuous loss can exceed the diffusion loss. Therefore the dependency of ${\Sigma_{con}^i}/{(1/\tau_{diff}^i)}$ for $i=e,p$ on the target density and magnetic field is shown in figures \ref{fig:electronmesh}-\ref{fig:protonmesh}.\\
\noindent In all variations that are meaningful for Cygnus X, which can be seen in figure \ref{fig:ChiAlpha}, the diffusion loss almost always exceeds other losses. In conclusion, it is crucial to consider the diffusion.\\
\noindent Retrospectively, this conclusion could give us some hints about the relation of the young supernova remnant $\gamma$-Cygni and the Cygnus Cocoon if we assume the same parameters for the Cocoon as for the whole Cygnus X. It must be mentioned that $\gamma$-Cygni may have delivered protons and electrons at TeV range approximately five kyrs ago (\cite{CygnusByFermi2}). So, our steady-state model is not appropriate enough for the Cygnus Cocoon but the whole Cygnus for the reason as mentioned above. However, assuming the same parameters for the Cygnus Cocoon as for the whole Cygnus X, we can give hints about the relation between $\gamma$-Cygni and Cygnus Cocoon.\\
%The Cygnus Cocoon lies between Cygnus OB2 association and $\gamma$-Cygni, which can be seen in figure \ref{fig:fermimapzoom2b}.  
According to \cite{CygnusByFermi2}, the condition for $\gamma$-Cygni to maintain as the only accelerator for the Cygnus Cocoon is that the dominant particle transport mechanism should be diffusion and the diffusion coefficient largely be similar to the coefficient in our Galaxy. In that case, the particles released by $\gamma$-Cygni could maintain CRs from the whole Cocoon. The diffusion-dominated scenario leads to an isotropic particle release from the young remnant. It is important to answer the question whether the particles from $\gamma$-Cygni could maintain the CRs in the Cygnus Cocoon due to diffusion transport mechanism. It can be answered by considering the mean free path and the average distance $d_{\gamma, Cyg}$ of $\gamma$-Cygni to the Cygnus Cocoon, which is pictured in figure \ref{fig:cyggal}.

\noindent Here, the solid blue line represents the mean free paths in Cygnus X, the red line the one in the Galaxy (\cite{D0diffCoeff}) and the dashed black line shows the distance of $\gamma$ Cygni from the Cocoon. %\footnote{The distance of $\gamma$ Cygni to the Cocoon was found by using Aladin V9.0 and considering the average distance of Cygnus X from the Earth.}
Nearly all CRs from $\gamma$ Cygni in direction to the Cocoon could reach the Cocoon if the mean free path is similar or larger than the mean free path of the Galaxy according to (\cite{CygnusByFermi2}) or at least longer than the distance of these objects to each other.\\
\noindent However, the mean free path is for electrons always smaller than the distance and than the value in the Galaxy. At energies $\gtrsim 10^5$ GeV, protons from $\gamma$-Cygni may be reasonable for the freshly accelerated CRs from the Cygnus Cocoon since the mean free path becomes larger than the distance between them. For lower energies indeed $\gamma$ Cygni is not favored to be the only injector of CRs.

\subsection{Parameters for Cygnus X}
\noindent Data from different observatories can be used to restrict the magnitude of the parameters on the one hand, and the number of free parameters on the other hand. In the following $\gamma$-ray data from Fermi, Argo and Milagro will be used. Additionally, the non-thermal radio data from the work of  \cite{thermalNonThermal1} and \cite{thermalNonThermal2} will be considered. When doing so, for example fitting methods can be used to find the normalization factor $q_0^{e,p}$ of the source rate function. The non-thermal radio data will be used to determine $q_0^e$ and $\gamma$-ray data to find $q_0^p$. Additionally, by considering emission in the energy range, which requires only electrons, the amount of injected electrons can be determined and thus the appearance of a leptonic process. Especially, the contribution of non-thermal Bremsstrahlung can be estimated. This correlation will be discussed in the results in chapter \ref{Results}. In order to constrain the appearance of leptonic processes, the differential fluxes for non-thermal radio emission and $\gamma$-rays from Cygnus X will be presented followed by the correlation between them.
In the following, a magnetic field strength of $B$=1 $\mu$G (\cite{Francis}) and a target density of $N_t$=70 cm$^{-3}$ (\cite{CygnusByFermi}) will be used. Using the brightness temperature spectral index of non-thermal data a spectral index of $\alpha$=2.6  (\cite{thermalNonThermal2,thermalNonThermal1}).
The Cygnus X region can be summarized by the following parameters:
\begin{table}[H]
	\centering
	\begin{tabular}{c||c}
		Parameters & Cygnus X \\
		\hline \hline Electron diffusion length [cm]& $l_{e}=2\times10^{16}$  \\ 
		Proton diffusion length [cm] &$l_{p}=l_{e}(\frac{m_{p}}{m_{e}})^{\beta}$  \\ 
		Diffusion index & $\beta=1/3$  \\ 
		Infrared photon density [eV/cm$^3$]& $U_{IR}= 5$ \\ 
		Advection velocity [km/s]& $v_{adv}=50$  \\ 
		Dust temperature  [K] & $T_{dust}= 25$  \\ 
		Distance to Cygnus X [kpc]& $d=1.4$\\ 
		Radius of Cygnus X [pc]&  $R= 77$ \\ 
		Radius of Cygnus X [$\deg$] & $\Theta_{CygX}=3.15$\\
		\hline\hline
	\end{tabular}
	\caption{Input parameters for the Cygnus X region.}
	\label{list1}
\end{table}

\section{Results with parameters from previous calculations (PC)}
\label{RFCM} 
\begin{figure}[H]
	\centering
	\subfigure{\includegraphics[width=1.\linewidth]{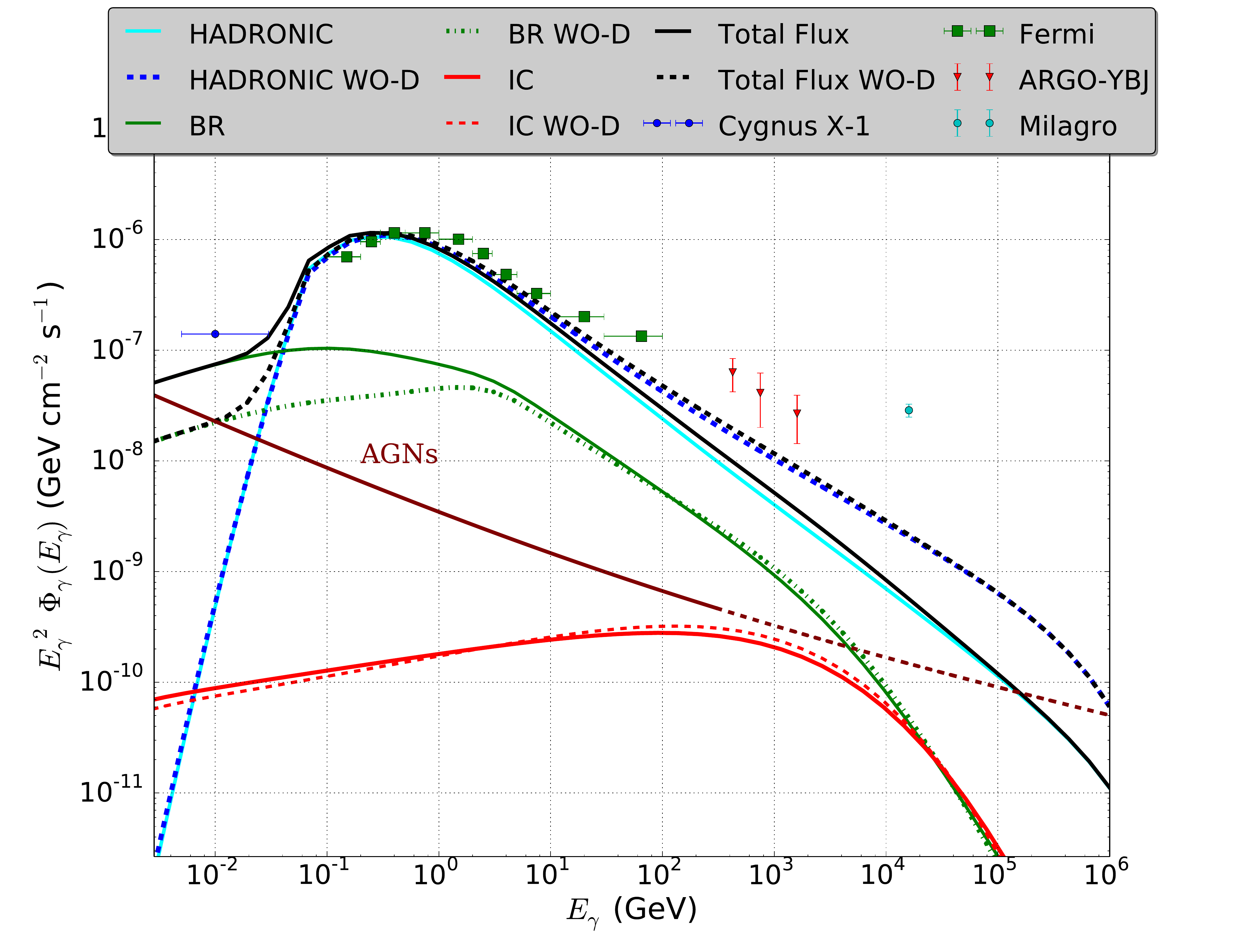}}
	\caption[Gamma-ray energy spectrum with old parameters]{$\gamma$-ray energy spectrum \textbf{with} and \textbf{without} consideration of diffusion loss; the source rate normalization factor $q_0$ was fitted on the observed $\gamma$-ray data. Additionally, the parameters PCs have been used.}
	
	\label{fig:aaa}
\end{figure}

\begin{figure}[H]
	\subfigure{\includegraphics[width=1.\linewidth]{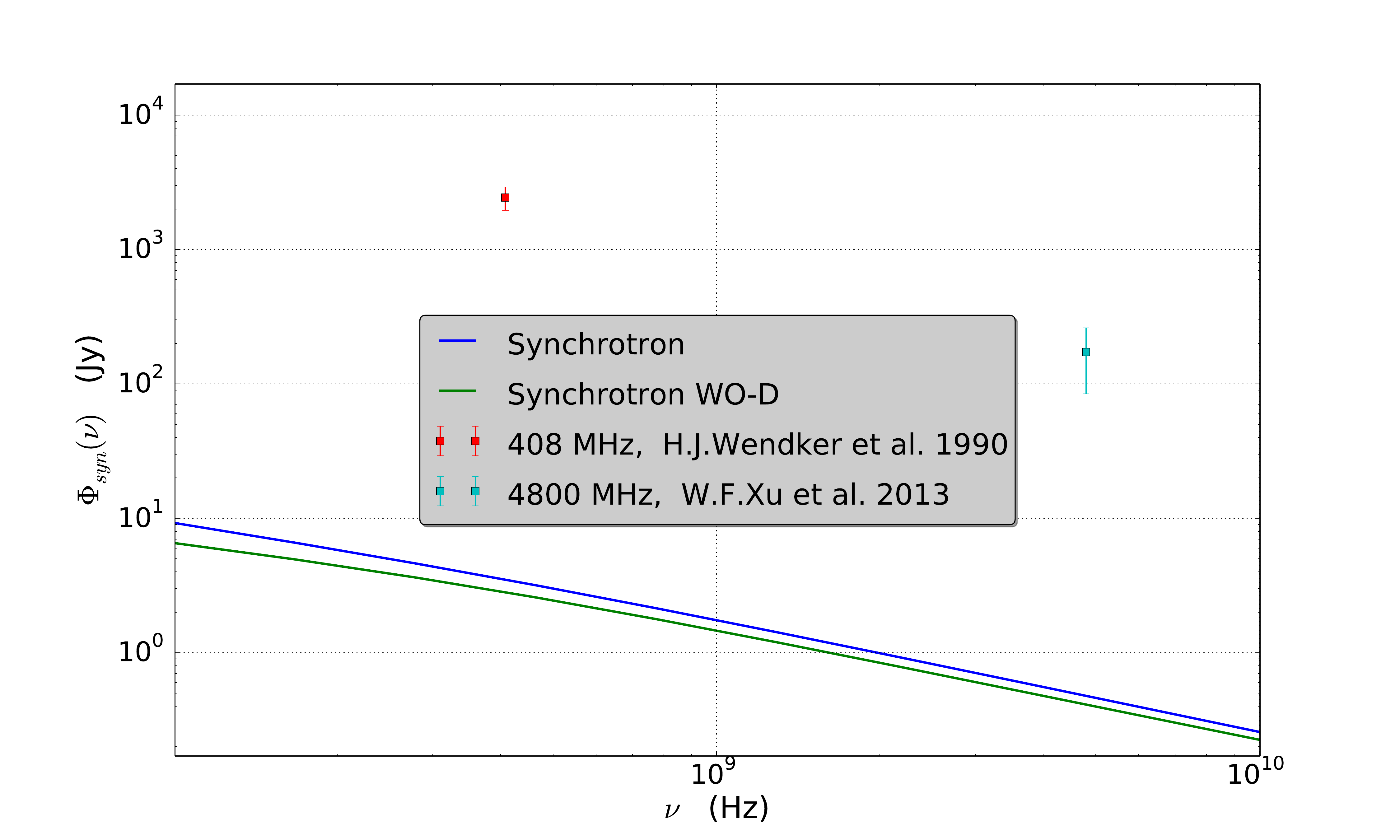}}
	\caption[Synchrotron spectrum with old parameters]{Synchrotron differential flux spectrum as a function of the frequency \textbf{with} and \textbf{without} consideration of diffusion loss; the source rate normalization factor $q_0$  was fitted on the observed gamma data. Here, the parameters from PCs have been.}
	\label{fig:aaa2}
\end{figure}

\noindent In the following, the non-thermal radio and $\gamma$-ray spectra will be presented by considering parameters from PCs (see figure \ref{fig:aaa} and \ref{fig:aaa2}). The consideration of the $\gamma$-ray spectrum is paramount as all relevant subatomic processes occur in it and the radio data assist in constraining the leptonic processes as it is based on one.
\noindent Here, "\textbf{IC}" denotes the contribution of the Inverse Compton process to the total differential $\gamma$-ray flux. In the same way "\textbf{BR}" denotes Bremsstrahlung and "\textbf{HADRONIC}" the hadronic pion production. The solid line represents the progression of the total differential $\gamma$-ray flux of this work. Besides, the results are presented without diffusion which is denoted as "WO-D", and with diffusion loss as diffusion is the main loss mechanism.
The diffuse fluxes observed by Fermi LAT (green cross), ARGO-YBJ (red triangle) and  Milagro (turquoise circle) are pictured. The Fermi $\gamma$-ray data points are taken from \cite{CygnusByFermi} and adapted for our region of interest. The same procedure is done for ARGO-YBJ from \cite{ARGOData} and Milagro from \cite{MilagroData}. From these diffuse fluxes the extragalactic point sources (AGNs) J2000.1+4212 and J2018.5+3851 are subtracted. It is not necessary to subtract galactic point sources as our model considers them by fitting the source rate normalization factor $q_0$ on the observed $\gamma$-ray flux.\\
Additionally, the flux from Cygnus X-1 measured by Comptel can be seen (blue circle).\\
Cygnus X-1 is a very well studied black hole and a front runner microquasar candidate in the Galaxy. It contains a dominant power-law component at 10 MeV. Consequently, it can be assumed that the flux from the whole Cygnus X region should be equal or higher.\\
Considering the constituents of the total flux, the progression at 10 MeV is of particular importance, since the total differential flux at 10 MeV is distinctly dominated by Bremsstrahlung which is caused by the leptonic process. The agreement with this data point and the consideration of the non-thermal radio data can point towards the real relevance of the diffusion loss mechanism. This is plausible as at the relatively low energy 10 MeV the entrapment of the electrons due to diffusion is more efficient than for higher energies (see figure \ref{fig:timescaleallE}) so that more Bremsstrahlung can be produced. 
Moreover, the real parameters such as the magnetic field and target density and the previous assumption regarding the diffusion of the particles can be investigated.\\
Comparing these results, there is no agreement between non-thermal radio and$\gamma$-ray data (see figure \ref{fig:aaa} and \ref{fig:aaa2}). The used parameters have been investigated with PCs which do not consider diffusion loss as this work does. The value of the magnetic field e.g. is averaged over the whole Galaxy and is therefore not accurate enough for our calculations. As the structure of Cygnus X is very complicated, and it has many constituents the target density also might be not accurate enough for our model. Moreover, for Cygnus X the non-thermal radio and $\gamma$-ray have not been correlated before. In this work, the assumed parameters do not lead to an agreement between data from these radiations. 
\subsection{Best-fit procedure}
\begin{figure}[H]
	\centering
	\includegraphics[width=1.1\linewidth]{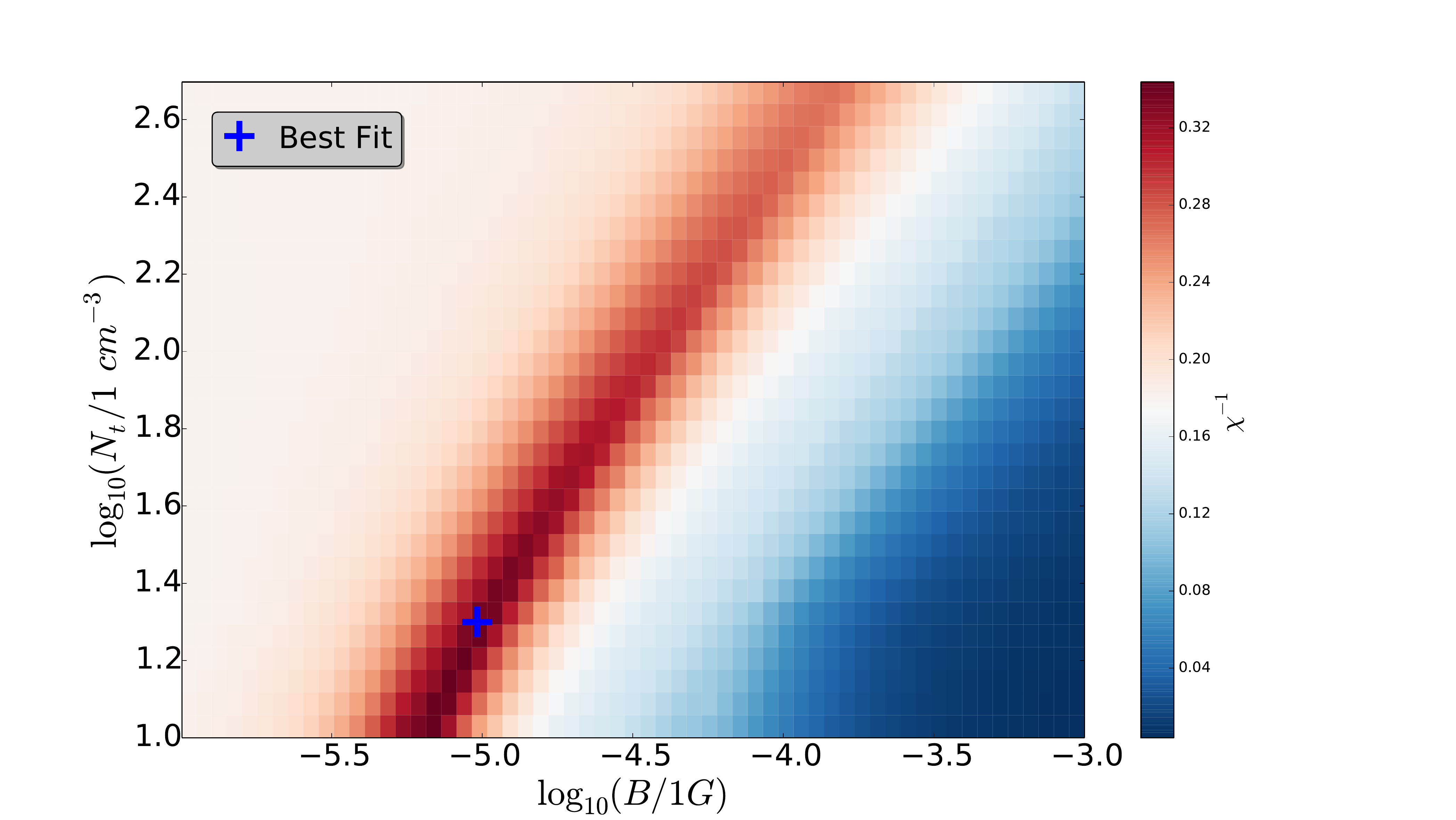}
	\caption[The total deviation]{The total deviation $\chi^2$ from the $\gamma$-ray and non-thermal radio data as a function of the magnetic field B and target density $N_t$. The best fit parameters are represented by "+". For each calculation of $\chi^2$ a best-adapted spectral index $\alpha\in[2.0-3.0]$ was also used.} 
	\label{fig:ChiAlpha}
\end{figure}

\begin{figure}[H]
	\centering
	\includegraphics[width=1.\linewidth]{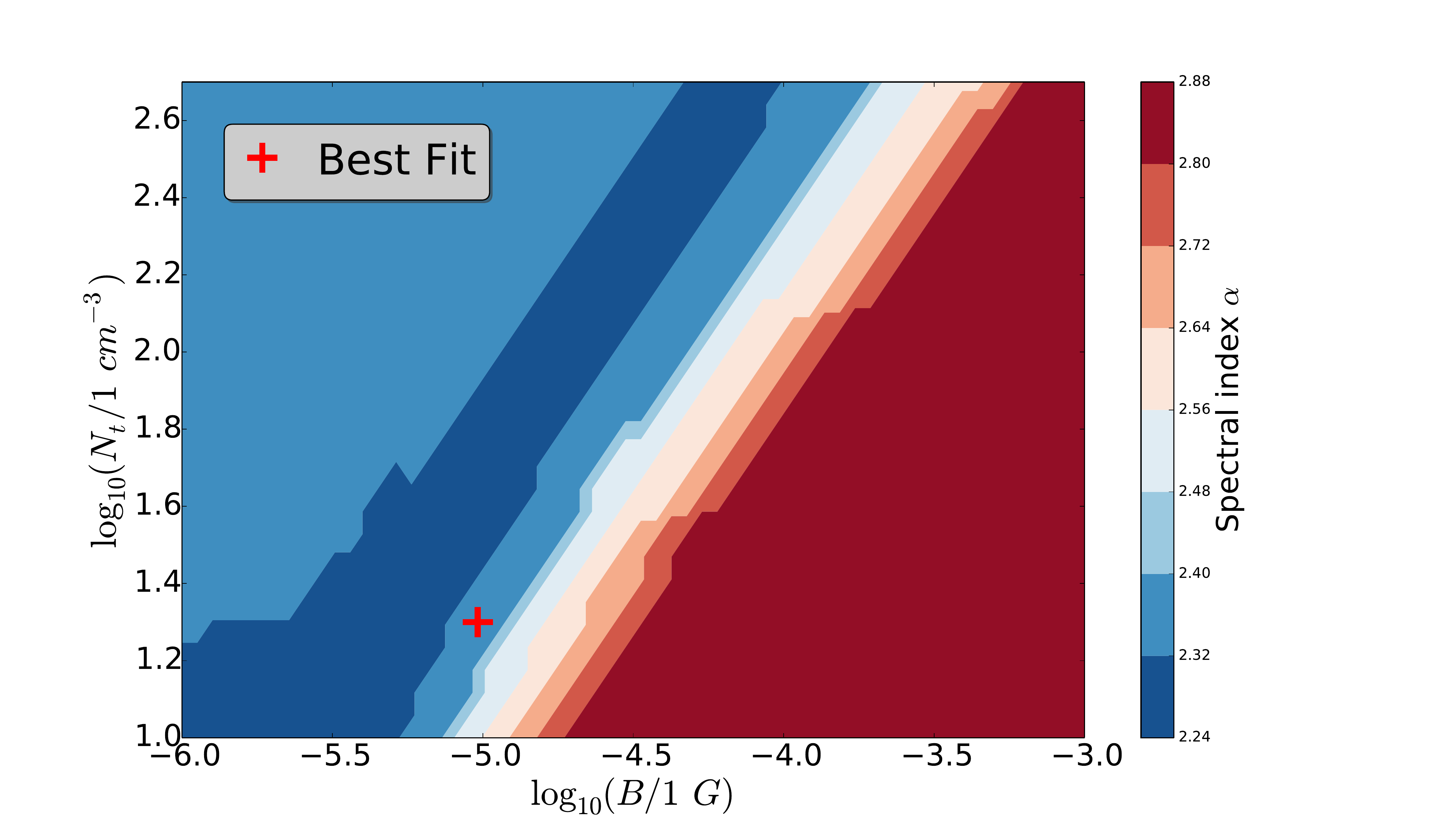}
	\caption[Best-adapted spectral index]{The best adapted-spectral index $\alpha$as a function of the magnetic field $B$ and target density $N_t$.}
	\label{fig:alpha}
\end{figure}

\begin{figure}[H]
	\centering
	\subfigure{\includegraphics[width=1.\linewidth]{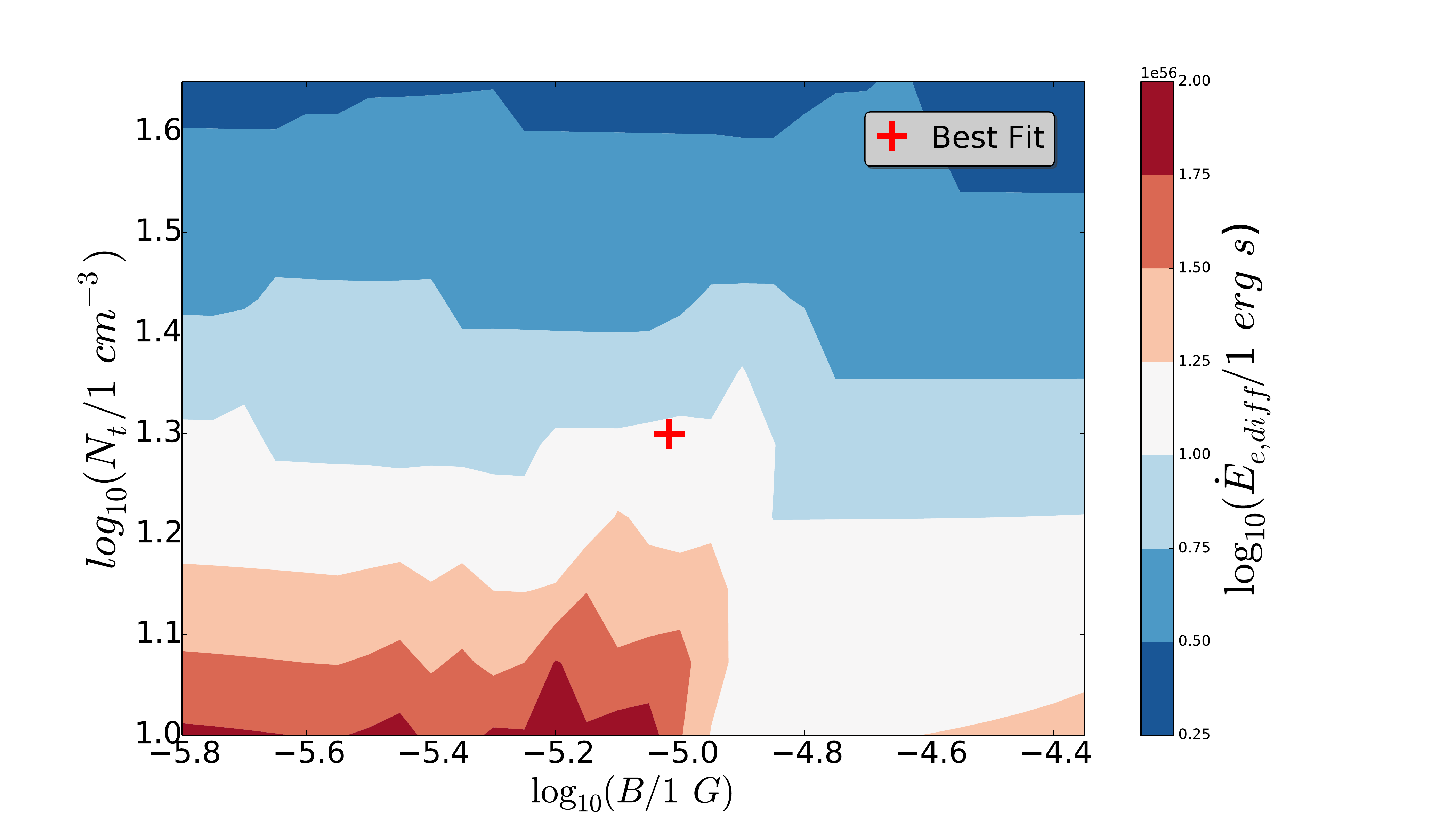}}
	\subfigure{\includegraphics[width=1.\linewidth]{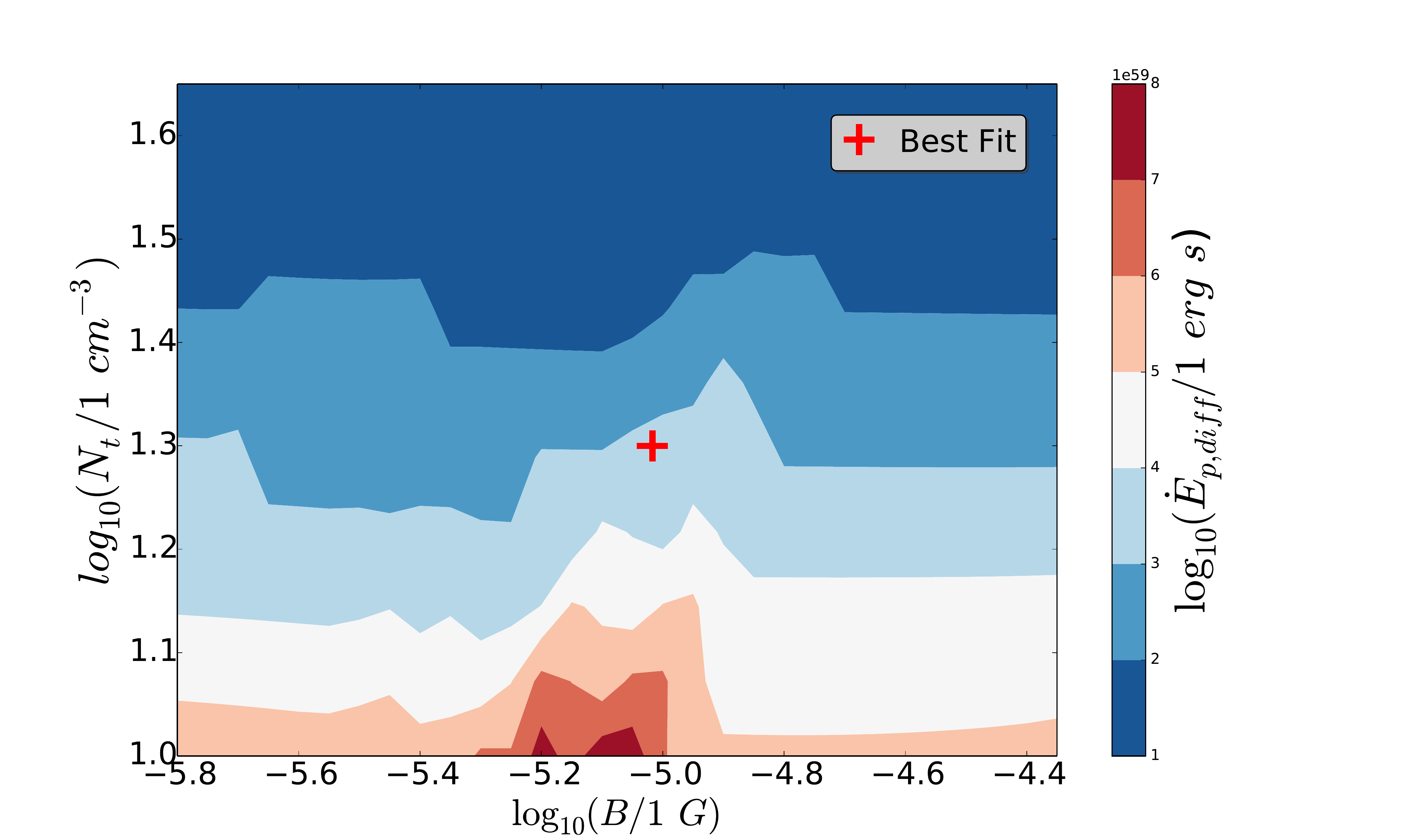}}
	\caption[Energy loss for protons and electrons]{Electron (upper) and proton (lower) energy loss $\dot{E}_{diff}$ in erg/s caused by diffusion as a function of the magnetic field $B$ and target density $N_t$.}
	\label{fig:energylosselectronzoom}
\end{figure}

\noindent Since the parameters used in the previous results do not seem to describe the Cygnus X sufficiently well, they must be changed.
On this basis, the total deviation in square $\chi^2$ of each data point to the theoretical flux is defined by the reduced $\chi^2$:
\begin{equation}
\chi^2= (\chi^2_{\gamma}+\chi^2_{\text{syn}})\, ,
\end{equation}
$\chi_{\gamma}$ denotes the deviation of the theoretical $\gamma$-ray flux from the $\gamma$-ray data over the degree of freedom. In the same regard, $\chi_{\text{syn}}$ describes the deviation of the theoretical synchrotron flux from the non-thermal radio data over the associated degree of freedom. They can be calculated with eq.(\ref{chi1}) and eq.(\ref{chi2}). The degree of freedom $F$ estimates the population standard deviation calculated from a sample. The degree of freedom is given by $F=n-1$, where $n$ is the size of the sample. In this work two samples have been used: Non-thermal radio and $\gamma$-ray data.
\begin{equation}
\begin{split}
\chi^2_{\gamma}&=\bigr[ \Phi_{\gamma,\text{obs}}(E_{\gamma})-\left( \Phi_{\gamma,IC}(E_{\gamma})+\Phi_{\gamma,Br}(E_{\gamma})\right.\\ &\left. +\Phi_{\gamma,had}(E_{\gamma})\right)\bigr] ^2 /\Delta\Phi_{\gamma,\text{obs}}(E_{\gamma}) ^2\cdot \frac{1}{F-1}
\end{split}
\label{chi1}
\end{equation} \begin{equation}
\chi^2_{\text{syn}}=\left( \dfrac{\Phi_{\text{syn,obs}}(\nu)-\Phi_{syn}(\nu) }{\Delta\Phi_{\text{syn,obs}}(\nu)}\right) ^2\cdot \frac{1}{F-1}\, ,
\label{chi2}
\end{equation}
Here, $\Delta\Phi_{\gamma,\text{obs}}(E_{\gamma})$ and $\Delta\Phi_{\text{syn,obs}}(\nu)$ denote the uncertainties which result from the measurements.\\
As discussed above it is a reasonable step to set the magnetic field and the target density as free parameter as the correlation of radio and gamma data proved problematic and the relation was strongly affected by these parameters. 
\noindent The variation of $B$ and $N_t$ can also change the spectral index $\alpha$. Furthermore, the diffusion of the particles causes a steepening of the spectrum and increases the flux produced by leptonic processes which are necessary to reach especially the data point at 10 MeV, as the entrapment of the particle due to diffusion is at the relatively low energy 10 MeV more efficient. The whole variation range of the three fitting parameters can be taken from table \ref{variation2}:
\begin{table}[H]
	\centering
	\begin{tabular}{c||c}
		
		Physical parameters & Variation range \\ 
		\hline \hline
		Magnetic field $B$ [G] & $[10^{-7};\ 10^{-4}]$  \\  
		Target density $N_t$ [cm$^{-3}$] & [$10^1;\ 10^{2.7}$]  \\ 
		Spectral index $\alpha$ & [$2.0;\ 3.0$]\\
		\hline \hline
	\end{tabular} 
	\caption{Variation range of magnetic field, target density and spectral index.}
	\label{variation2}
\end{table}
\noindent The deviation can be illustrated graphically, such that the influence through the whole range and the best-fit parameters may be seen easily. Figure \ref{fig:ChiAlpha} shows $\chi^2$ as a function of the target density $N_t$ and the magnetic field $B$.
\noindent The dark red area shows a stronger agreement between theoretical and experimental fluxes than the other colors.\\
These two parameters, in particular, appear to be markedly different than anticipated. The magnetic field is larger than the previously assumed value by a factor of nearly one order of magnitude, and the target density is smaller by a factor of 3.6. Considering figure \ref{fig:ChiAlpha}, the magnetic field is not supposed to be smaller than 3 $\mu$G or larger than 100 $\mu$G. The target density has an upper limit of 300 cm$^{-3}$. The best-fit point is represented by a "+" sign.\\
%\noindent The total deviation for a smaller variation range is pictured in figure \ref{fig:ChiAlpha225-245}.
The lowest $\chi^2$ provides a magnetic field strength of $B=8.9\times10^{-6}$ and a target density of $N_t=19.4$ cm$^{-3}$.\\
In order to find a reliable spectral index, the best-adapted spectral index is presented as a function of the magnetic field and target density in figure \ref{fig:alpha}.
\noindent If one considers the range from figure \ref{fig:ChiAlpha}, where the smallest $\chi^2$ was found, it may be asserted that $\alpha=2.32-2.4$ is the best value for Cygnus X.
However, the smaller range of variation and the calculations in this work show that the best-fit is obtained for $\alpha=2.37$. %Thus overall, it shows that the spectral index from \cite{thermalNonThermal1} and \cite{thermalNonThermal2} is reliable, which confirms the results in this work.\\
\\
%Since H. J. Wendker et al.1990 and W. F. Xu et al. 2013 have found the same value for $\alpha=2.6$, the variation in this work with considering diffusion loss should not be significant, as it does not consider diffusion loss and is similar to other conventional models. In the case of considering diffusion loss the spectral index appear to be 2.5 which does not differ greatly from the previous results.\\  \\
In addition, the energy loss in erg/s can be investigated by:
\begin{equation}
\dot{{E}}^i_j=\frac{4}{3}\pi R^3m_i c^2\int_{\gamma_{0}^i}^{\infty}d\gamma\gamma\frac{n_i(\gamma)}{\tau_j}\, ,
\end{equation}
whereby the index $i$ refers to the quantity of an electron or proton, $\tau$ the loss timescale and  $j$ the loss mechanism.
Since diffusion is the dominant loss mechanism, $\dot{{E}}^e_{diff}$ and $\dot{{E}}^p_{diff}$ as a function of the magnetic field and target density respectively will be presented in figure \ref{fig:energylosselectronzoom}. Here, for each data point, an adapted spectral index $\alpha$ and the source rate normalization factor $q_0^p$ have been used. The result is a calculation of the energy loss with different spectral indices and $q_0$, which lead to a smallest total deviation $\chi^2$.
\noindent In Cygnus X the energy loss yields $\dot{{E}}^e_{diff}=1.08\times10^{56}$ erg/s and $\dot{{E}}^p_{diff}=3.39\times10^{59}$ erg/s.
This implies that the proton is more prone to be lost due to diffusion than electrons. This behavior is comprehensible in view of figure \ref{fig:timescaleleelectron} and also eq.(\ref{sigmaDiff}), which show that $\lambda_p>\lambda_e$ and thus $\tau^p_{diff}<\tau^e_{diff}$.\newpage
\section{Results with best-fit parameters}
\label{Results}
\begin{figure}[H]
	\centering
	\includegraphics[width=0.85\linewidth]{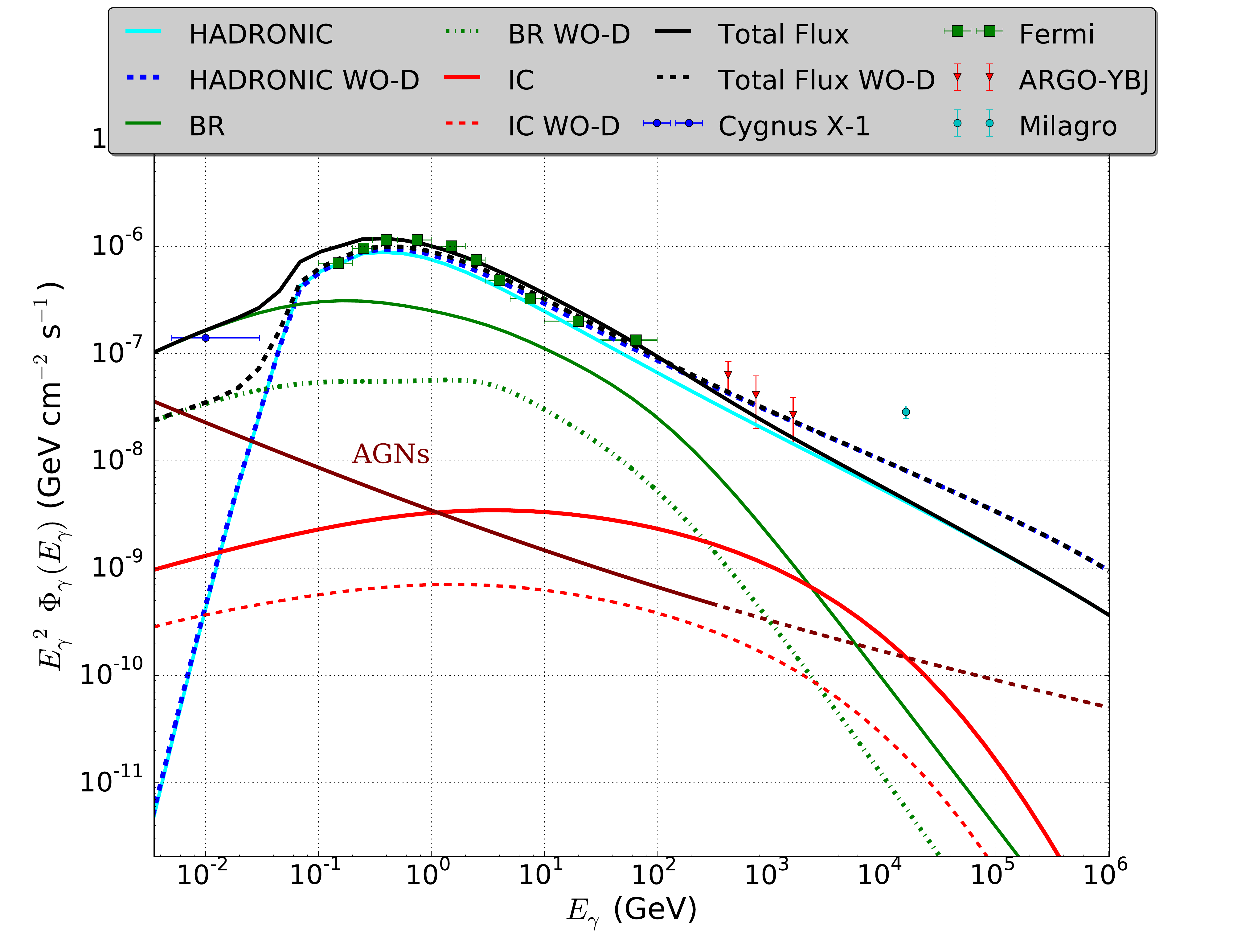}
	\caption[Gamma-ray energy spectrum with new parameters]{$\gamma$-ray energy spectrum \textbf{with} and \textbf{without} consideration of diffusion loss; the source rate normalization factor $q_0$ was fitted on the observed $\gamma$-ray data. Additionally, the new and better-adapted parameters from list \ref{list2} were used.}
	\label{fig:a252b24e-5nt23q0p236e-23}
\end{figure}

\begin{figure}[H]
	\centering
	\includegraphics[width=0.85\linewidth]{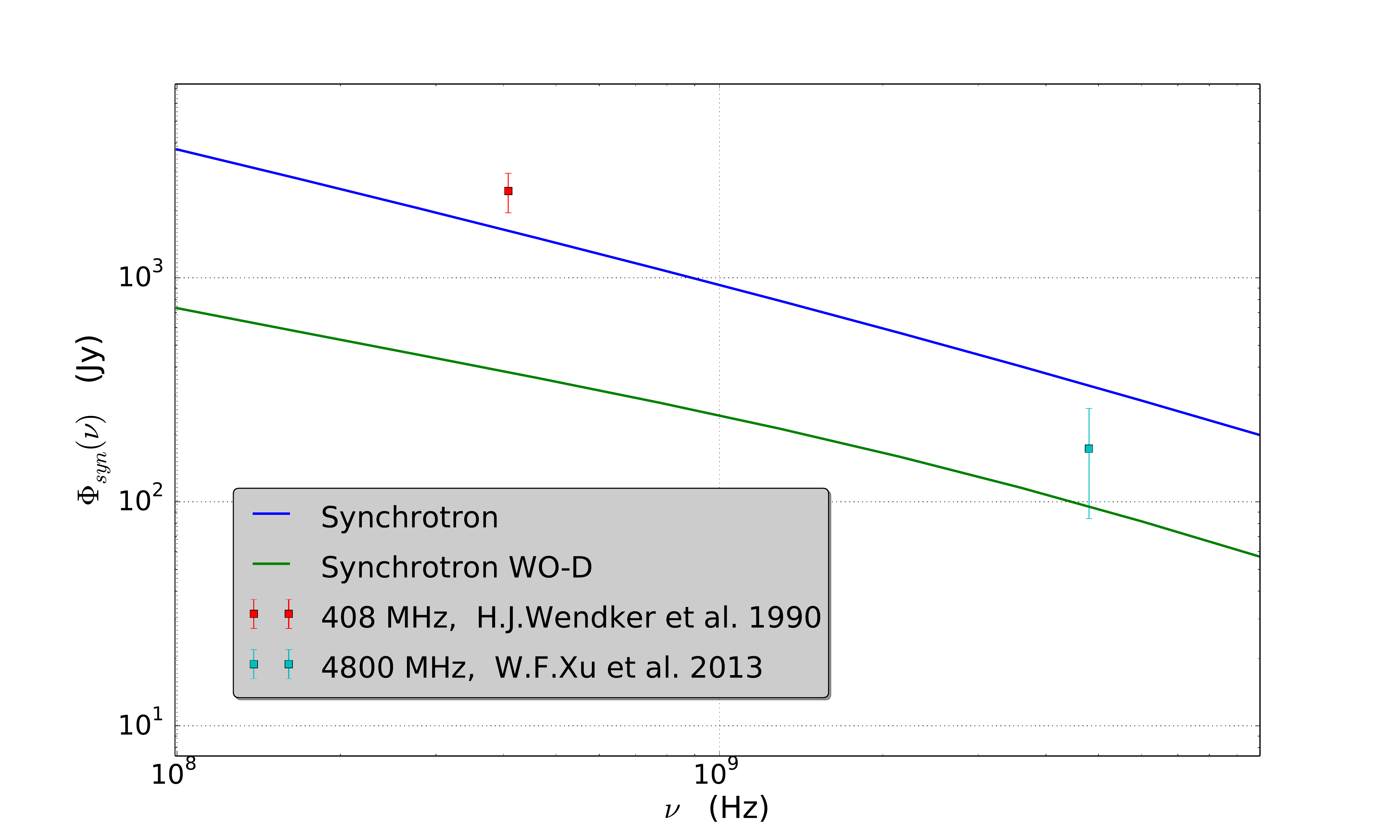}
	\caption[Synchrotron spectrum with new parameters]{Synchrotron differential flux spectrum as a function of the frequency \textbf{with} and \textbf{without} consideration of diffusion loss; the source rate normalization factor $q_0$  was fitted on the observed gamma data. Here, the new and better-adapted parameters from list \ref{list2} were used.}
	\label{fig:2a252b24e-5nt23q0p236e-23}
\end{figure}

\begin{figure}[H]
	\centering
	\subfigure{\includegraphics[width=0.8\linewidth]{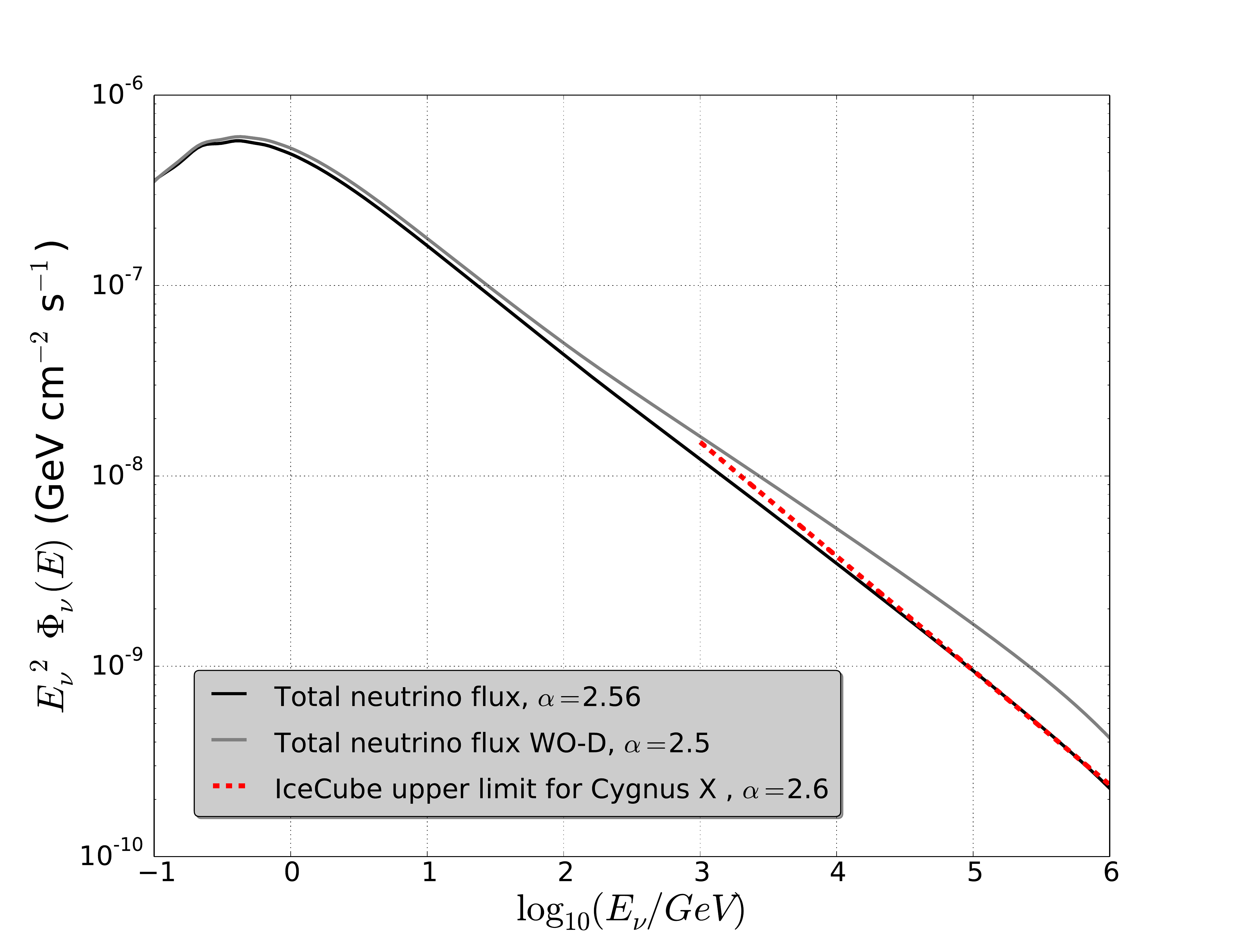}}
	\caption[Differential neutrino flux considering new parameters with Diffusion]{Differential neutrino flux considering new parameters \textbf{with} and \textbf{without} diffusion loss as function of the energy in GeV and IceCube upper limit calculated for Cygnus X (\cite{CygLimitDis}).}
	\label{fig:neutrinoNewPar}
\end{figure}

\begin{figure}[H]
	\centering
	\subfigure{\includegraphics[width=0.8\linewidth]{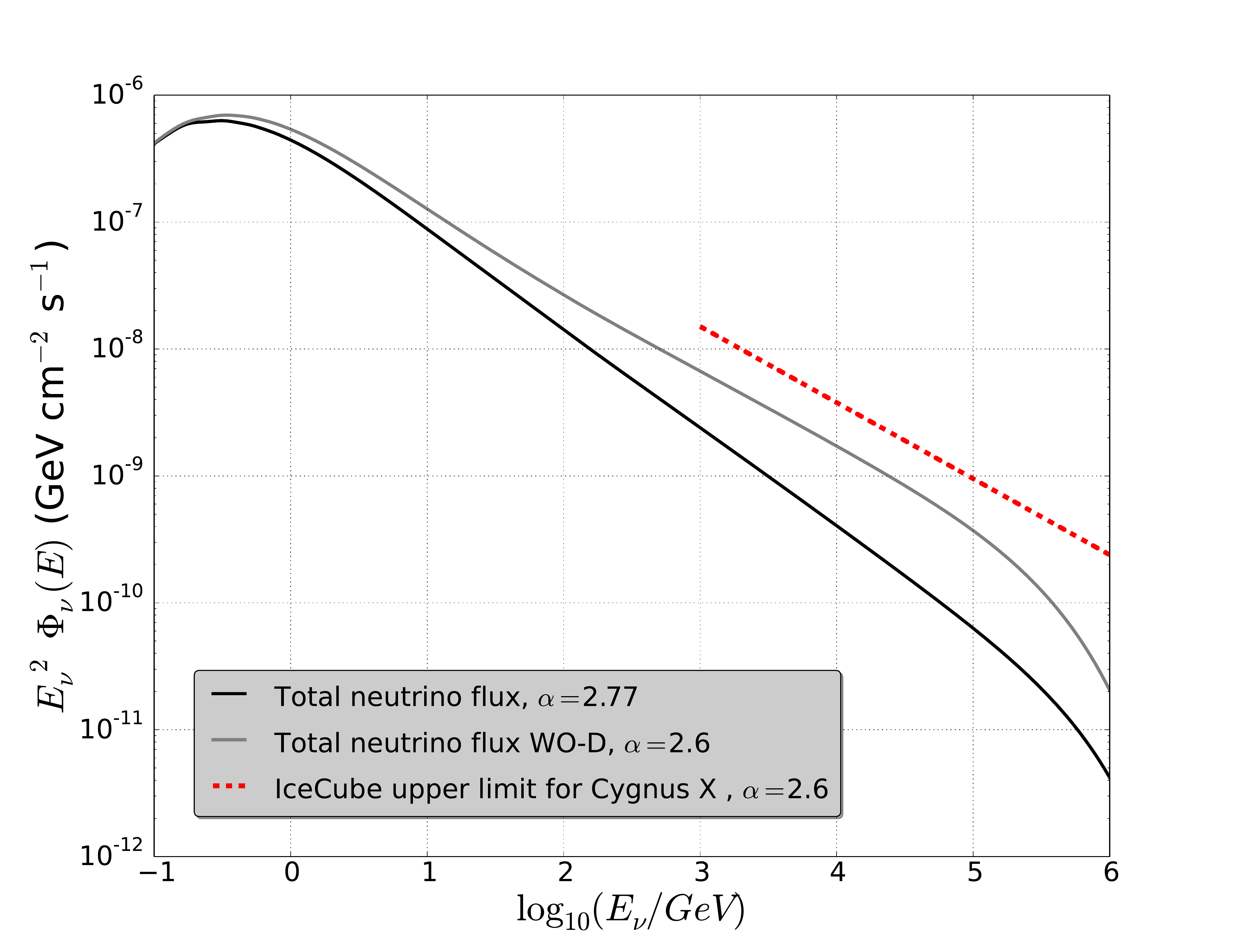}}
	\caption[Neutrino spectrum with old parameters]{Differential neutrino flux considering parameters from PCs \textbf{with} and \textbf{without} diffusion loss as function of the energy in GeV and IceCube upper limit calculated for Cygnus  X (\cite{CygLimitDis}).}
	\label{fig:neutrinoNewPar2}
\end{figure}

\noindent The best procedure leads to the following parameters
\begin{table}[H]
	\centering
	\begin{tabular}{c||c}
		Parameters & Cygnus X \\
		\hline \hline Magnetic field strength [G]& $B=8.9\times10^{-6}$ \\ 
		Proton target density [cm$^{-3}$]& $N_t=19.4\ $ \\ 
		Spectral index& $\alpha=2.37$ \\ 
		Proton source rate \\ normalization factor  [cm$^{-3}$ s$^{-1}$]& $q_0^p=9.8\times10^{-22}$ \\ 
		Electron source rate \\ normalization factor cm$^{-3}$ s$^{-1}$]& $q_0^e\approx 172\times q_0^p$ \\ 
		\hline\hline
	\end{tabular}
	\caption{Best fit parameters for the Cygnus X region.}
	\label{list2}
\end{table}
\noindent The new target density provides the depth of Cygnus X in the spiral arm, as the the column density in Cygnus X is known. Again considering the work of \cite{CygnusByFermi} a column density of $C_{H_I}=70$ cm$^{-2}$ is obtained. Putting the target and column density in a simple relation, the neutral gas distribution over a depth of $d_t$ is obtained:
\begin{equation}
d_t=C_{H_I}/N_t\approx3\times10^{20}\ \rm cm\approx116\ \rm pc
\end{equation}
Furthermore, the relation between $q_e(\gamma)$ and  $q_p(\gamma)$ leads to an injection rate of protons 172 times greater than that of electrons, i.e.  $q_p(\gamma)\approx 172\times q_e(\gamma)$. Finally, the $\gamma$-ray and non-thermal radio spectrum can be presented in figure  \ref{fig:a252b24e-5nt23q0p236e-23} and \ref{fig:2a252b24e-5nt23q0p236e-23}, respectively.\\
\noindent The $\gamma$-ray spectrum shows that for high energies ($>40$ MeV) the dominant constituent of the total differential $\gamma$-ray flux is caused by the hadronic pion production. The Bremsstrahlung represents the second largest component of the total flux. This process is also the most significant element of $\gamma$-ray flux at lower energies ($<40$ MeV). The Inverse Compton differential flux is for the most part in the background.
\noindent The new results for $\gamma$-rays show an adamant agreement between ARGO-YBJ and Fermi data. The condition for the flux at 10 MeV is also fulfilled. In contrast, the Milagro differential flux is less than circa four times greater than the differential flux calculated in this work. This leads to the suspicion that Milagro may be overestimating the flux in Cygnus X. Indeed, Daniele Gaggero et al. 2015 (\cite{AliPaper}) have shown for the Galactic diffuse emission that a conventional representative model with the properties found by Fermi data fails to reproduce the large flux measured by Milagro, meaning the overestimation of Milagro is still an open issue. As our diffuse flux can not explain the Milagro flux, a point source, i.e. might me responsible 
at these high energies. In a recent paper \cite{TovaPaper} have evaluated the radiation in the Cygnus region resulting from the propagation of the average cosmic ray flux in the Galaxy. The result depends on the spectrum assumed but, in any case, cannot accommodate the observed photon emission, especially at TeV energy and above as measured by Milagro and HAWC. The conclusion is that accelerators in the region, most likely in the Cocoon, must be responsible for the high-energy flux. This is consistent with our modeling of the Cygnus.\\
Nevertheless, a natural explanation for the "Milagro anomaly" in our Galaxy has been found by considering the radial dependence for diffusion coefficient spectral index $\beta$  and the advective wind. However, this fact does not hold great relevance for the present work, since Cygnus X is small in comparison to the Galaxy.
The prediction that the agreement between theoretically and experimentally determined fluxes deteriorates when examining high-energy particles may therefore be validated.
Also, the new parameter leads to a much stronger agreement for the non-thermal radio data than before.\\
In the same way, we identified the theoretical spectra without considering diffusion of the particles. The agreement between non-thermal radio and $\gamma$-ray data is worse than with considering diffusion and the conditions at 10 MeV is not fulfilled. This also shows us that diffusion is very meaningful for Cygnus X.\\
Finally, the neutrino differential flux spectrum considering the new parameters is presented in figure \ref{fig:neutrinoNewPar}. As a comparison, the spectrum considering parameters from PCs is pictured in figure \ref{fig:neutrinoNewPar2}. The limits in these figures are normalized with an $E^{-2.6}$ spectrum of Cygnus X from \cite{CygLimitDis}.
%The limits in figure \ref{fig:neutrinoNewPar}  are normalized with an $E^{-2.6}$ spectrum of Cygnus X from \cite{CygLimitDis} and adapted for a  $E^{-2.56}$ and $E^{-2.5}$ (WO-D) spectrum. The limit in figure  \ref{fig:neutrinoNewPar2} are adapted for an $E^{-2.8}$ and $E^{-2.62}$ (WO-D) spectrum. Here, it was assumed that the limits are similar at 100 TeV, as at this energy IceCube has the highest  sensitivity and the spectral index does not differ much from each other.\\
The first spectrum predicts a flux which coincides with the limit of IceCube at very high energies ($>$50 TeV). As IceCube has the highest sensitivity at 100 TeV and the spectral index of the predicted flux and the limit does not differ much from each other,
a significance measurement by IceCube or IceCube-Gen2 may be soon possible.\\
According to the model without diffusion loss (WO-D) the predicted flux is above the sensitivity of IceCube.
The neutrino differential flux at 100 TeV when considering diffusion loss is here almost 2.3 times smaller than the model without diffusion loss.\\
The second spectrum is worth measurable than the first one. Even the flux from WO-D  which is not realistic, is not measurable.
Overall, the $\gamma$-ray and non-thermal radio spectra provides a very strong agreement suggesting that the used parameters, transport mechanism, and model indeed seem to describe Cygnus X in an accurate way.

\section{Conclusion}
\label{Summary}
\noindent
%Summary of methods
In this paper, the leptonic and hadronic cosmic ray transport in the Cygnus is modeled with a leaky box model that takes into account energy loss processes via radiation and interaction as well as advective and diffusive transport, assuming the homogeneous injection of cosmic rays into Cygnus X. The solution of the transport equation is based on a semi-analytical approach, while the resulting radiation processes are fit to the broadband multiwavelength spectrum with a statistical procedure.
Because of the complex structure of Cygnus X and the miscellaneous processes which take place there, this work distinguishes between them and considers all relevant scenarios as far as possible, determining the best-fit scenario to draw conclusions about the possible dominance of a diffuse cosmic ray sea for the radiation signatures. 
Moreover, the relation between the electron and proton source rate is based on the quasi-neutrality of the plasma and depends primarily on the spectral index $\alpha$. Therefore, one aim of this work is to find a reliable spectral index which considers all relevant transport and cooling mechanisms, whereby a consistent spectral index of $\alpha=2.37$ is found. If electrons and protons are not injected with the same spectral index, or if they are injected with different minimal energies, the value for $q_e/q_p$ may change significantly. As there is no concrete evidence for differences in the spectrum the standard number was used.\\
Concerning the parameters of the interstellar medium, in our model, the radio flux in contrains a combination of the number of electrons and the magnetic field. The MeV emission can only be explained by bremsstrahlung losses, fixing the number of electrons in combination with the target densities. Highe-energy $\gamma$-rays then further constrain the electron and magnetic field strength via the inverse Compton process, while the proton number in combination with the target density are relevant for those $\gamma$-rays coming from $\pi^{0}-$ decays.
\\
At first, $\gamma$-ray and synchrotron spectra have been presented with those parameters that were used in early fits to the high-energy data. For this case, it could be shown that either the progression of the predicted $\gamma$-ray flux or radio flux do not satisfy the observed data.
%Moreover, the used parameters do not lead to a correlation between them. If these parameters are reliable, then an explanation of the discrepancy between the $\gamma$ and radio data in a physical regard should be possible. Indeed there are several explanations:
%\begin{itemize}
%	\item The high non-thermal radio flux may be caused by a background radiation.
%	\item Since Milagro is also overestimating the theoretical flux, the radio flux and the Milagro flux could have the same origin, i.e.\ a source which is not identified yet.  
%\end{itemize}
%However, a further explanation is that the parameters are not reliable.
In a second step, a best-fit procedure has been performed to find stable parameters. This fit leads to the following conclusinos:
\begin{enumerate}
	\item The fit parameters lead to an adamant agreement between predicted fluxes and the data measured by Fermi, ARGO-YBJ also to the non-thermal radio data, while the $10$~TeV data of Milagro are not well-fit.
	\item It can be shown that diffusion dominates the loss processes in Cygnus X and is important to consider in the transport equation: By considering the flux of Bremsstrahlung at 10 MeV, which is the dominant radiation process at this energy, a mean free path of $2\cdot10^{16}\cdot\gamma^{1/3}$ cm is found. The energy loss due to diffusion is also investigated. The protons in Cygnus X lose nearly $3.6\times10^3$ more energy due to diffusion than electrons.\\
	If we transfer this information to the Cygnus Cocoon, we can assert that at energies $ \lesssim10^5$ GeV the freshly accelerated protons in the Cygnus Cocoon have unlikely their origin only in $\gamma$-Cygni. 
	\item
	The condition for the flux at 10 MeV particularly fixes the contribution from bremsstrahlung, which is the only process that is able to contribute in such a high amount at these energies. This in term implies strong constraints for the main input parameters of the bremsstrahlung process, i.e.\ the differential electron number and the target column depth.
	\item The Milagro differential flux at $\sim 10$~TeV energies is about a factor of four times higher than the differential flux calculated in this work. These results show that a diffuse, homogeneous component could be responsible for the multiwavelength spectrum up to TeV energies, but a further component is necessary to explain the data at 10 TeV energies. Such an additional component could be a localized and/or short term accelerator within the region like $\gamma$-Cygni.
	%The high flux is suggested to be caused by an accelerator which is most likely located in the Cygnus Cocoon.\\ %The poorer agreement confirms the expectation that the used model is not appropriate for very high energies, as figure \ref{fig:fermimapzoom12} clarifies that Cygnus X demands a spatially inhomogeneous distribution.\\
	\item The new parameters provide a neutrino flux which approaches the sensitivity of IceCube at very high energies ($>$50 TeV). Considering that the difference between the spectral index of the flux and limit of IceCube for Cygnus X is less than 0.05, the coincidence is surely valid at high energies. In the future, the flux sensitivity of IceCube will be improved, so that the sensitivity for Cygnus X will suffice to measure the neutrino flux within the next decade.\\
	Additionally,  considering the relation between column and target density, our results indicate that the depth of the neutral gas $d_t$ of Cygnus X should be close to $d_t$=116 pc.
\end{enumerate}
Overall, despite its complexity, the present work investigates Cygnus X in a fundamental way, so as to reveal certain information about the transport mechanism, injection of cosmic rays and possible sources for acceleration.
\newpage
\section*{Acknowledgments}
\noindent We would like to thank the people of the "Theoretische Physik IV" of Ruhr-Universi\"at Bochum and the IceCube Collaboration for stimulating discussions and Rosa-Luxemburg-Foundation for supporting financially. We would like to express our profound gratitude to Steven Young Eulig, Mike Kroll, James Doing, Fabian Bos, Sebastian Sch\"oneberg, Lukas Merten,  Donglian Xu, Tova M. Yoast-Hull, Markus Ahlers, Paolo Desiati but especially to Ali Kheirandish.
Moreover, we are grateful for the comments of Dominik Bomans and Reinhard Schlickeiser.\\
We further acknowledge the support from the MERCUR project St-2014-0040 (RAPP Center) and the BMBF, FZ 05A14PC1.

	%\newpage
	\pagestyle{empty}
	\section*{References}
	\bibliographystyle{apsrev}
	\bibliography{literatur}
	
\end{document}